\documentclass[preprint]{elsarticle}
\usepackage{amssymb,amsmath,amsthm,latexsym}
\usepackage{mathrsfs}
\usepackage{a4wide}
\usepackage{collectbox}
\usepackage[usenames,dvipsnames]{color}
\usepackage[utf8]{inputenc}
\usepackage[T1]{fontenc}
\usepackage{dsfont}
\usepackage{tikz}
\usepackage{float}
\usepackage[mathscr]{euscript}
\usepackage{mathtools}
\usepackage{hyperref}
\usepackage{verbatim}
\biboptions{sort&compress}
\usepackage{amsmath,graphicx,tipa,amssymb}

\usepackage[section]{placeins}

\usepackage{hyperref}
\hypersetup{
    colorlinks = true,
    citecolor=blue,
    linkcolor=blue,
    linktocpage=true,
    linkbordercolor = {white},
}
\usepackage{color}

\usepackage{xurl}
\usepackage{lineno}
\modulolinenumbers[5]

\newcommand{\PP}{ \mathbb{P} }
\newcommand{\EE}{ \mathbb{E} }

\newcommand{\dt}{\delta t}
\newcommand{\rar}{\rightarrow}
\newcommand{\bv}[1]{\boldsymbol{\mathbf{#1}}}
\newcommand{\pa}{\partial}
\newcommand{\la}{\langle}
\newcommand{\ra}{\rangle}

\DeclareMathOperator{\lgt}{logit}
\DeclareMathOperator{\ept}{expit}

\begin{document}

\begin{frontmatter}

\title{Public efforts to reduce disease transmission implied from a spatial game}
\tnotetext[mytitlenote]{Fully documented templates are available in the elsarticle package on \href{http://www.ctan.org/tex-archive/macros/latex/contrib/elsarticle}{CTAN}.}

\author[mymainaddress]{James Burridge\corref{mycorrespondingauthor}}
\cortext[mycorrespondingauthor]{Corresponding author}
\ead{james.burridge@port.ac.uk}

\author[mymainaddress]{Micha\l \ Gnacik}
\ead{michal.gnacik@port.ac.uk}

\address[mymainaddress]{School of Mathematics and Physics, Lion Gate Building, Lion Terrace, University of Portsmouth, Portsmouth, United Kingdom}

\begin{abstract}
One approach to understand people's efforts to reduce disease transmission, is to consider the effect of behaviour on case rates. In this paper we present a spatial infection-reducing  game model of public behaviour, formally equivalent to a Hopfield neural network coupled to SIRS disease dynamics. Behavioural game parameters can be precisely calibrated to geographical time series of Covid-19 active case numbers, giving an implied spatial history of behaviour. This is used to investigate the effects of government intervention, quantify behaviour area by area, and measure the effect of wealth on behaviour. We also demonstrate how a delay in people's perception of risk levels can induce behavioural instability, and oscillations in infection rates. 
\end{abstract}


\end{frontmatter}

%\linenumbers

\section{Introduction}

The Covid-19 pandemic has demonstrated the importance of human behaviour in controlling disease transmission \cite{mcg20,fer06,thu20,kha21}.  In the absence of pharmaceutical interventions, governments around the world have imposed rules or recommended changes in behaviour to reduce infection rates. We refer collectively to these behavioural adaptations, which include social distancing, wearing face masks, reducing mobility, interactions and contacts, as \textit{infection-reducing behaviour}. In many cases these behavioural changes have reduced the reproduction number of the disease to less than one, and prevented the breakdown of healthcare systems \cite{thu20,kha21}. Although there are self-interested reasons for individuals to  change their behaviour, including personal safety and avoiding legal penalties, their actions can also be driven by a sense of civic duty \cite{dur21}: an altruistic desire to help one's fellow people. The costs and benefits to society as a whole of behavioural measures to reduce disease spread has been investigated \cite{thu20_2, row20}, and must also be considered by decision makers and individuals.

For many people the costs of complying with the restrictions are high, both economically and socially. For example, many lower-income individuals face a difficult choice between earning essential income and following the rules, which can reduce social distancing  \cite{ken21}. For every person, the decision to comply with government advice, or to take other steps to reduce disease transmission, must balance competing costs and benefits.  These may include personal safety, feelings of duty or guilt (which depend on the behaviour of others), economic losses, and the importance of social contact. Situations in which people are faced with behavioural choices, with payoffs determined in part by the behaviour of others, may be described mathematically using the theory of games \cite{Neu53,hof98,May82}. In this context, the collective non-pharmaceutical effort to control the virus is analogous to the \textit{public goods game}. In this game, every player must choose whether to contribute to a public pot. For us, contributing to the pot is equivalent to following governmental restrictions and cooperating with the collective effort to control disease. The pot's value is multiplied by a factor greater than one (but less than the number of players), and then shared between everyone. In the context of disease, the benefits of cooperation are reduced deaths and keeping health services running. Although the total payoff to the group is maximized if all members contribute, the rational agent does best by not contributing, and the only Nash equilibrium \cite{Nash51} is for no players to contribute. This situation is referred to as \textit{the tragedy of the commons} \cite{har68}. In reality, community cooperation is ubiquitous, leading game theory researchers to try to understand its origin. Experiments suggest that people are \textit{conditional cooperators}, meaning they will tend to cooperate at least as much as the community norm \cite{fehr18}. In other words, there is a psychological cost associated with cooperating less than others. Field studies show that populations with larger fractions of conditional cooperators better maintain common resources \cite{rus10}. Certain conditions, such as stable group composition, partner matching, the ability to punish freeloaders (provided the punisher does not benefit) and reward cooperators, may enhance group cooperation. These observations are borne out by simple spatial game models, which  show that territoriality may promote regional cooperation, and that cooperation is further enhanced by the ability to punish \cite{bra02}. In the current pandemic, mobility data suggest that social distancing can spread spatially \cite{coo20}, indicating the importance of spatial modelling.

Game theory has been applied to understand the interaction between disease and behaviour in the non-spatial setting, including its effect on vaccination rates \cite{bau04}, wearing masks \cite{krt21} (where the payoff gain depends on the difference between the perceived payoff of a mask wearer and the payoff for risking infection), and social distancing \cite{rel10,bha19,ara21} (where game payoffs balance infection risks against social distancing costs). These latter models couple game dynamics to the classical compartmental susceptible-infected-recovered (SIR) disease model \cite{ker27} by allowing transmission rates to depend on levels of social distancing. Behavioural effects have been incorporated in other non-spatial models, for example, by incorporating a social distancing term into the infection rate in SIR \cite{lux21}, by directly proposing an analytical relation between transmission rates and the disease state variables \cite{gou21,mwa20} or by varying parameters in compartmental models to perform scenario analyses, or estimate transmission rates from infection data \cite{gio20}. The effects of social distancing on disease spread have also been studied using spatial models, either using networks \cite{sil19,mah20} or spatially varying fields \cite{vru20,giu20,tso21} (discrete or continuous). In simpler models, spatial dynamics are explored by exogenously specifying transmission rates or distancing behaviour, and then solving for various hypothetical scenarios \cite{tso21,mah20}. More sophisticated spatial models explicitly model behaviour dynamics, using techniques borrowed from physics such as density functional theory \cite{vru20} and multiplex networks \cite{sil19}. These methods have been used to explore hypothetical scenarios, but stop short of fitting to real spatial data. A simpler model, which does not explicitly model the dynamics of disease or behaviour, has been used to infer the importance of transmission within and between Italian provinces from publicly available spatio-temporal disease data \cite{giu20}.

\subsection*{Outline of the paper}
In this paper we present a spatial game model (section \ref{sec:model}), coupled to disease dynamics (section \ref{ssec:dis}), which can be calibrated to high-resolution spatio-temporal case data (section \ref{sec:calib}). Due to the severity of the  current pandemic, large financial and human resources have been diverted to measure and record infections with broad coverage and high spatio-temporal resolution. Using simple assumptions about the effects of infection-reducing behaviour on disease spread, we calibrate our behaviour model to this data (section \ref{sec:calib}). Our spatial game, played by conditional cooperators, leads to a  logistic-linear model \cite{was03,has09} of behavioural response (section \ref{ssec:behdyn}). The collective behaviour of spatial cells in our model is formally equivalent to a Hopfield neural network \cite{hop82,hop84} in which connection weights are determined by geographical proximity and behavioural parameters. By fitting these parameters to spatial disease data we can infer how behaviour has evolved during the pandemic in the UK, showing how it is affected by government lockdowns and infection levels.  This allows us to measure how well government actions (e.g., the tier system) have worked and if the public have complied with them (section \ref{sssec:tier}). We also find significant statistical correlation between wealth and implied behaviour in London (section \ref{sec:behwealth}). Our approach offers the possibility of understanding the relationship between game parameters, and measurable quantities like mobility and retail activity. It may also be useful for devising optimal control measures. Finally we explore how infection rates and delayed information affect the stability of the behavioural dynamics (section \ref{sec:delay}).

\section{The model}

\label{sec:model}
\subsection{Behaviour dynamics}
\label{ssec:behdyn}
People can reduce disease transmission rates in many ways \cite{fer06}, but for simplicity we assume that each $i$-th individual in the population is described by a state or ``strategy'' variable $S_i(t) \in \{0,1\}$ where $S_i(t)=1$ at time $t$ if they are acting to reduce the chance of catching or transmitting the disease. We refer to  people in this state as \textit{cooperators}. Their behaviour may often be in response to governmental restrictions or guidance, or it may be a matter of personal initiative. People who are not infection-reducing are in state $S_i(t)=0$ and we call them \textit{defectors}. We divide space into cells and define the average cooperation rate in the cell with centroid $\bv{r}$ as
\begin{equation}
u(\bv{r},t) = \frac{1}{N(\bv{r})} \sum_{i \in \la \bv{r} \ra} S_i(t)
\end{equation} 
where  $\la \bv{r} \ra$ is the set of individuals whose homes are in cell $\bv{r}$, and $N(\bv{r})=|\la \bv{r} \ra|$, is
the cell population. 

Now we describe the interactions between cells. To model them we introduce the \textit{interaction matrix} $W$, with entries given by
\begin{equation}\label{eqn: exposure_mat}
W(\bv{r}, \bv{r}^{\prime}) = \frac{N(\bv{r}')}{\mathcal{N}(\bv{r})} \exp\left\{- \frac{|\bv{r} - \bv{r}^{\prime}|^2}{2\sigma^2}\right\},
\end{equation}
where $\sigma$ is the \textit{interaction range}---the typical distance between the home locations of people who observe or interact with each other---and $\mathcal{N}(\bv{r})$ is a normalizing constant which makes $W$ a stochastic matrix \cite{gri20}, that is, $\sum_{\bv{r}'} W(\bv{r},\bv{r}') =1$ for all $\bv{r}$. The element $W(\bv{r},\bv{r}')$ is the fraction of observations or interactions made by individuals in cell $\bv{r}$, which are of individuals in cell $\bv{r}'$. For simplicity, in this paper we model interactions generated by \textit{physical proximity} (which could allow disease transmission) in the same way as \textit{observations} (which facilitate behavioural copying). We also keep the interaction matrix constant in time, meaning that the typical interaction range is not affected by the overall level of cooperation. However, when we model disease transmission, interaction matrix elements are adjusted to account for the collective effects of infection-reducing behaviour (see section \ref{ssec:dis}). We set $ \sigma = 10$km %(as $ 6.8 \text{miles} \approx 10 \text{km}$ rounded to first significant figure) 
as this is the approximate average trip length in England in 2019 \cite{dist20}.  
%The normalization ensures that the total exposure is one. 
The definition of the interaction matrix may be made more sophisticated in future work, for example by allowing $\sigma$ to depend on time and place, or to be different for observations than for interactions, if sufficient mobility or interaction data is available for calibration.

Using the interaction matrix, the behaviour observed by individuals in cell $\bv{r}$ is
\begin{equation} \label{eqn: mean_beh}
\bar{u}(\bv{r},t) = \sum_{\bv{r}'} W(\bv{r},\bv{r}') u(\bv{r}',t).
\end{equation}
We write $P_i(s)$ for the payoff to player $i$ of using strategy $s \in \{0,1\}$. The payoff for cooperation includes the safety benefits of collective action minus the personal costs of reducing disease transmission. We assume the simplest linear form consistent with these assumptions
\begin{equation}
P_i(1) = \alpha_i \bar{u} - k_i,
\end{equation}
where $\alpha_i$ is a risk reduction factor, and $k_i$ is a cost. Defectors also benefit from the collective effort, but they don't pay the cost. However, assuming they are conditional cooperators \cite{fehr18,fehr00}, feelings of guilt increase with the community norm  cooperation level, so
\begin{equation}
P_i(0) = (\alpha_i-\gamma_i) \bar{u}
\end{equation}
where $\gamma_i$ is the \textit{guilt factor}. The sign of the payoff difference
\begin{equation}
\Delta_i = P_i(1)-P_i(0) = \gamma_i \bar{u}-k_i
\end{equation}
determines the optimal strategy at any given time. Intuitively, if the  cooperation levels are low, people feel little guilt in failing to contribute, and are less willing to pay the costs of complying with the restrictions. If  cooperation rates are high, they will be willing to pay the cost of performing their civic duty \cite{fehr18}. For simplicity we assumed that cooperators and defectors get the same risk reduction, and that, excluding costs, the payoffs of the two strategies are the same when $\bar{u}=0$. Relaxing these assumptions shifts the cost and guilt factor, but does not change the linear form of $\Delta_i$. In either case there is a critical ratio
\begin{equation}
r_i = \frac{k_i}{\gamma_i}
\end{equation} 
for which cooperation is favourable provided $\bar{u}>r_i$.

We now assume that individuals spontaneously switch to the optimal state (either $0$ or $1$ depending on the sign of $\Delta_i$ for each individual $i$) with probability $\tau^{-1}$ per unit time, where $\tau$ is a \textit{time constant}.  As $\dt \rar 0$, we have the following Markov process \cite{gri20}
\begin{align}
\PP(S_i(t+\dt)=s | S_i(t)=s) &= 1 - \frac{\dt}{\tau} \left(1 - s \mathbf{1}_{\bar{u}>r_i} - (1-s) \mathbf{1}_{\bar{u}<r_i}\right) + o(\dt)\\
\PP(S_i(t+\dt)=1-s | S_i(t)=s) & = \frac{\dt}{\tau} \left( (1-s) \mathbf{1}_{\bar{u}>r_i} + s \mathbf{1}_{\bar{u}<r_i}\right) + o(\dt),
\end{align}
where 
\begin{equation*}
    \mathbf{1}_{A} = \begin{cases}
    1 & \text{ if event $A$ occurs} \\
    0 & \text{ otherwise}
    \end{cases}.
\end{equation*}
The time constant, $\tau$, gives the expected time between updates. Defining $\delta S_i(t) = S_i(t+\dt)- S_i(t)$ we have
\begin{equation}
\EE\left[ \frac{\delta S_i(t)}{\dt} \middle \vert S_i(t)=s \right] = \frac{1}{\tau} \left( \mathbf{1}_{\bar{u}>r_i} - s \right) + \frac{o(\dt)}{\dt}.
\label{dS}
\end{equation}
Now consider cell $\bv{r}$ which we assume contains a large number of individuals. Let $F(r)$ be the distribution of critical ratios in the cell, then averaging (\ref{dS}) over the cell, and taking the limit $\dt \rar 0$, we have
\begin{align}
\EE_t \left[ \dot{u}(\bv{r},t)\right] & = \frac{1}{\tau} \left(\int \mathbf{1}_{\bar{u}>r} dF(r)  - u(\bv{r},t)\right)  \\
&= \frac{1}{\tau} \left(F(\bar{u}(\bv{r},t))  - u(\bv{r},t)\right)
\end{align}
where $\EE_t$ is the expectation taken conditional on the state of the system at time $t$. We  assume cells are sufficiently large that stochastic fluctuations ($O(N^{-1/2})$) can be neglected, and  $\dot{u}(\bv{r},t)$ approximated with the above expectation (median cell population in our simulations is $\tilde{N} = 281120$). In this case 
\begin{equation}
\dot{u}(\bv{r},t) = \frac{1}{\tau} \left(F(\bar{u}(\bv{r},t))  - u(\bv{r},t)\right).
\label{eqn:learn}
\end{equation}
We take (\ref{eqn:learn}) as the definition of our model.  In the absence of measurements of $F$, a natural choice is to take it to be normal. However, the model is more tractable if we choose $F$ to be the logistic distribution.
Recall that the logistic (cumulative) distribution function  with location $c \in \mathbb{R}$ and scale $s>0$ is given by 
\begin{equation}
F(x) = \frac{1}{1+ e^{-\frac{x-c}{s}}}= \frac{1}{2} + \frac{1}{2} \tanh\left(\frac{x-c}{2s} \right).
\label{eqn:logistic_dist}
\end{equation}
The normal distribution can be approximated very closely using the logistic, with the difference in cumulatives never exceeding 0.01  \cite{sav06}. Taking the variance of the logistic distribution to be
\begin{equation}
\sigma^2 = \frac{\pi^2}{3 \beta^2}
\end{equation}
or equivalently the scale parameter to be $s = \beta^{-1}$
where $\beta>0$, we obtain
\begin{equation}
\lgt(F(\bar{u}(\bv{r},t)) = \beta(\bar{u}(\bv{r},t)-c)
\label{eqn:beh}
\end{equation}
%$$F(\bar{u}(\bv{r},t)) = \frac{1}{1+\exp{\{-\beta(\bar{u}(\bv{r},t)-c)}\}}= \frac{1}{2} + \frac{1}{2}\tanh{\left(\frac{\bar{u}(\bv{r},t)-c}{2\beta} \right)}  $$
where $c = N^{-1}\sum_{i} r_i$ is the mean critical ratio, and the ``log odds'' function is defined $\lgt(u) = \ln(u/(1-u))$. We refer to $F(\bar{u})$ as the \textit{response function}, because it represents the   cooperation rate toward which the cell evolves.

The constant $\beta^{-1}$ measures the variation in thresholds amongst the population, which may come from differences between individuals, or from variations in the behavioural constants of individual players at different times. These latter variations are equivalent to introducing an element of irrationality in decision making, where the decision function $\mathbf{1}_{\bar{u}>r_i}$ is replaced with a smoothed step. In this case $\beta$ may be viewed as an \textit{inverse decision temperature}. The origin of this terminology may be understood by writing the spatially homogeneous form of our model (\ref{eqn:learn}) in terms of the variables $m=2u-1$, $h=1-2c$, with time in units of $\tau$ 
\begin{equation}
    \dot{m} = \tanh \left( \frac{\beta(m + h)}{4}  \right) - m.
\end{equation}
This is the mean field equation for the magnetization, $m \in [-1,1]$, of the Ising--Glauber model \cite{kra10,bra94} in an applied field of strength $h$  with inverse thermodynamic temperature $\beta/4$. The model describes the alignment of magnetic spins in a ferromagnetic material. For sufficiently low temperature and provided $c \in [0,1]$, two fixed points emerge in this model corresponding to two alternative bulk alignments within the material \cite{che05,bra94}. In our case these aligned states correspond to universal cooperation or universal defection (the population is capable of ``tragedy'', if enough people defect). The coarse grained spatial behaviour of the Ising model is described by the time dependent Ginzburg--Landau (TDGL) equation \cite{kra10,bra94}, which allows different parts of the spatial domain to align in different directions, with curvature driven interfaces between. Our spatial model is analogous to the TDGL equation, so could in principle develop such interfaces, provided the decision temperature was low enough. However, in the period of time we have studied (section \ref{sec:sim}) we do not find evidence of such subcritical coarsening behaviour.

The form (\ref{eqn:learn}) of our behavioural dynamics is about the simplest possible model of a population in which individuals learn from the behaviour of others. In fact it is formally equivalent to the continuous Hopfield neural network \cite{hu03,hop82,hop84,mac03}, allowing for self-connection. According to the Hopfield model, the activity, $x_i$, of the $i$th neuron is driven by a weighted sum of the activities of the neurons whose outputs feed into it. This sum is known as an \textit{activation}, defined $a_i=\sum_j w_{ij} x_j$, where $w_{ij}$ are connection weights. In the Hopfield model, the response of the neuron to this activation is governed by the equation
\begin{equation}
\dot{x}_i = \frac{1}{\tau} \left(f(a_i)-x_i \right)
\end{equation}
where $f$ is known as the \textit{activation function} and $\tau$ is a time constant \cite{mac03}. The equivalence to our model may be observed if we identify our cells as neurons, our exposure matrix elements $W(\bv{r},\bv{r}')$ as connection weights, and our response function as the activation function. The only difference to the Hopfield model is that $W(\bv{r},\bv{r})>0$, meaning that neurons can connect to themselves. In principle one could treat (\ref{eqn:learn})  as the fundamental definition of the model, and then select a response function $F$ based on general considerations rather than microscopic derivation. For example, $F$ must map $[0,1] \rar [0,1]$. We also expect it to be an increasing function, because humans tend to copy the behaviour of others. It might also contain terms which are independent of others' behaviour. From this perspective, the logistic response is amongst the simplest possible choices, and an alternative parameterization which emphasises this, and is useful for analytical work (section \ref{sec:delay}), is 
\begin{equation}
\lgt(F(\bar{u},t)) = -\beta_0 + \beta_1 \bar{u}(\bv{r},t).
\label{eqn:beh2}
\end{equation}
Here $\beta_0$ represents behavioural responses which are independent of others', whereas $\beta_1$ measures the strength of behavioural coupling.

%For practical simulation purposes, we use the discrete model
%\begin{equation}
%u(\bv{r},t+1) = \left( 1-\frac{1}{\tau} \right) u(\bv{r},t) + \frac{1}{\tau} F(\bar{u}(\bv{r},t)).
%\end{equation}
%We will measure time in days, in which case $\tau$ may be viewed as the typical number of days individuals take to respond to changes in behavioural parameters and the behaviour of others. 

\subsection{Coupled disease-behaviour model}
\label{ssec:dis}
Recent experience has shown that behaviour change can dramatically reduce the rate at which disease spreads \cite{mcg20,kha21}. It also seems likely that infection rates may affect behaviour in return (we explore this possibility in section \ref{sec:delay}). To explore these relationships we first couple our behaviour model to a spatial Susceptible-Infected-Recovered-Susceptible (SIRS) model \cite{ker27}, which provides a simplified description of Covid-19 dynamics \cite{com20}. We let $S(\bv{r},t), I(\bv{r},t)$ and $R(\bv{r},t)$ be fractions of individuals who live in cell $\bv{r}$ who are susceptible, infected or recovered at time $t$. Infections spread via contact between susceptible and infected individuals. 

To capture the effects of  infection-reducing behaviour on the spread of disease between cells, we write disease transmission rates in terms of the interaction matrix $W$ and behaviour $u(\bv{r},t)$. We first define the \textit{infection-exposure matrix}
\begin{equation}
\widetilde{W}(\bv{r},\bv{r}') = (1-u(\bv{r},t))W(\bv{r},\bv{r}') (1-u(\bv{r}',t)).
\end{equation}
Here we have assumed that, due to the reciprocal nature of contacts, the exposure of individuals in $\bv{r}$ to $\bv{r}'$, and vice-versa, is proportional to $(1-u(\bv{r},t))(1-u(\bv{r}',t))$. This is the probability that two randomly selected individuals, one from $\bv{r}$ and one from $\bv{r}'$ are both defecting at time $t$. Infection reduction can be achieved both by staying at home, and by taking the appropriate precautions when in public, and for simplicity our model treats these two possibilities as equivalent. As emphasized earlier (in section \ref{ssec:behdyn}), we do not allow the form of the interaction matrix used to model behavioural observation to depend on behaviour. That is, in the current version of our model, we are assuming that the typical distance between the home locations of people who observe one another remains constant. 

\begin{figure}[h!]
	\centering
\includegraphics[scale=0.6]{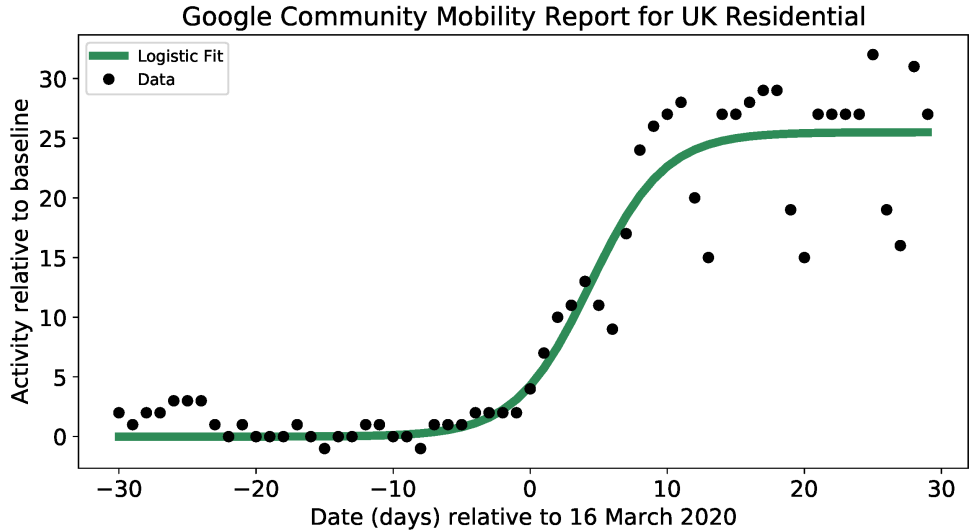} 
	\caption{Data points show Google community mobility reports \cite{goo20} for residential activity 30 days either side of 16 March 2020. Green curve shows least squares fit of logistic curve (\ref{eqn:logistic}) to data, yielding values $a=25.3$, $\tau = 2.7$ and $t_0=4.3$. }
	\label{fig:google_res}
\end{figure}

Given the infection-exposure matrix, the infection probability per unit time experienced by susceptible members of cell $\bv{r}$ is then the following functional of $I$ and $u$ (with these quantities viewed as functions of $\bv{r}$)
\begin{equation}
\hat{I}_{\bv{r}} [I,u] = \sum_{\bv{r}'} \widetilde{W}(\bv{r},\bv{r}') I(\bv{r}',t).
\label{eqn:Ihat}
\end{equation}
Using measure (\ref{eqn:Ihat}) of exposure to infection, the probability that a susceptible individual in cell $\bv{r}$ is infected during time interval $\dt$ is $\epsilon \hat{I}_{\bv{r}}[I,u]\dt$, where $\epsilon$ is the baseline infection rate, in the absence of any cooperation. Assuming that in time $\dt$, infected individuals recover with probability $p \dt$ and recovered individuals lose immunity with probability $\xi \dt$, then  in the deterministic limit ($N(\bv{r}) \rar \infty$), disease and behaviour evolve as follows
\begin{align}
\label{eqn:du}
\dot{u}(\bv{r},t) &= \frac{1}{\tau} \left( F(\bar{u}(\bv{r},t)) - u(\bv{r},t) \right),\\
\dot{S}(\bv{r},t) &= \xi R(\bv{r},t)- \epsilon \hat{I}_{\bv{r}}[I,u] S(\bv{r},t), \\
\dot{I}(\bv{r},t) &= \epsilon \hat{I}_{\bv{r}}[I,u] S(\bv{r},t) - p I(\bv{r},t),\\
\dot{R}(\bv{r},t) &= p I(\bv{r},t) - \xi R(\bv{r},t).
\label{eqn:dR}
\end{align} 
We will refer to these equations as the Behavioural SIRS model, or BSIRS for short. 
Our estimates of disease (Covid-19) and other model parameters (excluding $\beta_0$ and $\beta_1$) are given in Table \ref{tab:par}. The basic reproduction number $\mathcal{R}_0$ gives the expected number of secondary cases per infection in a completely susceptible population. In the case of SIRS dynamics (excluding the behaviour) we have 
\begin{equation}
\mathcal{R}_0 = \frac{\epsilon}{p}.
\end{equation} 
Recent estimates of median $\mathcal{R}_0$ for English regions \cite{liu21} have mean $3.5$. UK government guidance requests a 10 day isolation period from the onset of symptoms, which we use as our estimate for mean period of infectivity, so $p=0.1$ and $\epsilon=0.35$. A recent serological study ($\approx 6000$ participants) suggests that immunity lasts for at least 5-7 months \cite{rip20} and based on this we set the rate of immunity loss to $\xi=0.07$.
\begin{table}%[H] add [H] placement to break table across pages
	 \centering
	\begin{tabular}{ccccccc}
		\hline
		Parameter &  $\mathcal{R}_0$ & $\epsilon$  & $p$ & $\xi$ & $\tau$ & $\sigma$ \\
		\hline
		Value & 3.5 & 0.35 & 0.1 & 0.007 & 3 & 10km \\
		\hline
	\end{tabular}
	\caption{\label{tab:par} Table of parameter values for the behavioural SIRS (BSIRS) model, assuming time is measured in days. Disease parameters $\mathcal{R}_0$, $\epsilon$,   $p$  and $\xi$ were estimated from \cite{liu21,rip20}. The time-constant $\tau$ was estimated from Google mobility reports (see Figure \ref{fig:google_res}) and the interaction range $\sigma$ is the approximate mean trip length in England in 2019 \cite{dist20}. }
\end{table}
We estimate the memory length using mobility reports \cite{goo20, mob20, mobR20}. For example, Google mobility reports use location history to determine activity levels in various classes of location, including parks, residential areas and workplaces. After UK government advice was issued on 16 March 2020 to cease all unnecessary social contact, residential activity increased approximately logistically before reaching a stable equilibrium level after $\approx 15$ days. Strongly publicised national government edicts of this kind would be represented within our modelling framework as a rapid exogenous change in parameters. The time scale over which the public responds to such edicts gives a measure of the time constant in our model, which we estimate by fitting a logistic curve
\begin{equation}
\label{eqn:logistic}
\ell(t) = \frac{a}{1+\exp \left(-\frac{t-t_0}{\tau} \right)}
\end{equation}
to the change in activity levels at the start of the first lockdown, using least squares, yielding $\tau \approx 3$. The function $\ell$ differs from the logistic distribution function (\ref{eqn:logistic_dist}) only by multiplicative factor, $a$. The  results of the fit are shown in Figure \ref{fig:google_res}.

\section{Results}
\label{sec:calib}

\subsection{Calibration methodology}

To perform simulations and calibrate to disease data, we assume that the model parameters in Table \ref{tab:par} are fixed, leaving two free behavioural parameters, $\beta$ and $c$, in each cell. To avoid over-parameterization, we assume that the decision temperature is spatially invariant, and allow the mean critical ratio to depend on position and time, subject to regularization. Therefore
\begin{equation}
\lgt(F(\bar{u}(\bv{r},t)) = \beta(\bar{u}(\bv{r},t)-c(\bv{r},t)).
\end{equation}
To calibrate $c(\bv{r},t)$ to case data, consider the 3-month period 29 September 2020 to 28 December 2020. Since the immunity lasts for approximately $5-7$ months and the simulation period is shorter, the SIR model would provide an adequate approximation to disease dynamics during this period. However, to allow for later flexibility, and since there is uncertainly about the distribution of reinfection times, we developed simulations to work in the most general setting so that parameters could later be adjusted. During our period of interest there is complete NHS Covid-19 case data for England \cite{covid_data}. Data from 2021 is excluded, because the mass vaccination program changed disease constants  ($\varepsilon$, $p$ and $\xi$). We divided our three month period into nine intervals, each of length ten days. We then determined the total number of active cases at interval boundaries, in each of the 151 \textit{upper tier local authorities} (UTLAs) in England. We assumed a 10 day infection duration, consistent with the 10 day self-isolation recommendation \cite{govgui}. The UTLAs form the cells of our model, and their populations are set to Office of National Statistics estimates from mid-2019 \cite{popest}. 

To allow direct comparison to data, in this section we let $S(\bv{r},t),I(\bv{r},t),R(\bv{r},t)$ represent \textit{absolute case numbers in each cell}. We also discretize the evolution equations (\ref{eqn:du})-(\ref{eqn:dR}) into one day time steps, yielding the simulation algorithm
\begin{align*}
u(\bv{r},t+1) &= \left(1 - \frac{1}{\tau}\right)u(\bv{r},t) +   \frac{1}{\tau}F(\bar{u}(\bv{r},t)),\\
S(\bv{r},t+1) &= \left( 1 - \frac{\epsilon}{N(\bv{r})} \hat{I}_{\bv{r}}[I,u]\right)S(\bv{r},t) + \xi R(\bv{r},t), \\
I(\bv{r},t+1) &= ( 1 - p )I(\bv{r},t) + \frac{\epsilon}{N(\bv{r})} \hat{I}_{\bv{r}}[I,u] S(\bv{r},t) ,\\
R(\bv{r},t+1) &= (1 - \xi) R(\bv{r},t) + p I(\bv{r},t).
\end{align*}
Normalized case rates (for which $S+I+R=1$ in every cell) are obtained by dividing simulated numbers by the corresponding cell populations $N(\bv{r})$. We perform simulations for a series of decision temperatures selected from the range $\beta \in [0.05,4]$. For each $\beta$ value we calibrate the critical ratios $c(\bv{r},t)$ so the case rates match the NHS data. When $\beta > 4$ the calibration results become noisy and the convergence is slow.

Let $t$ be the time in days from the start of our disease dataset, which consists of $n=10$ arrays each containing $m=151$ entries representing the number of active Covid-19 infections in the corresponding UTLA, at $t=0,10,20, \ldots, 90$. Let $I^{\text{data}}_k(\bv{r})$ denote the number of active cases taken from the data in cell (UTLA) $\bv{r}$ at $t=10k$ where $k \in \{0,1,\ldots, 9\}$. We assume that critical ratios are constant over each 10 day interval, and write $c_k(\bv{r})$ for the critical ratio in cell $\bv{r}$ during the time interval ending at $t=10k$  where $k \in \{1,2,\ldots, 9\}$.  We define the vector
\begin{equation}
\bv{c}_k = (c_k(\bv{r}_1), c_k(\bv{r}_2), \ldots , c_k(\bv{r}_m)).
\end{equation}
We also define 
\begin{equation}
I_k(\bv{r}) = I(\bv{r},10k)
\end{equation}
for $k \in \{0,1,\ldots, 9\}$, where $I(\bv{r},t)$ is computed from simulations with initial condition $I(\bv{r},0)=I^{\text{data}}_0(\bv{r})$. To calibrate the values of critical ratio vectors $\bv{c}_k$ we aim to minimise the following mean squared error with a regularisation term, added to minimise the fluctuations in behaviour between neighbours
\begin{equation}
MSE_{\text{reg}} = \frac{1}{(n-1) \cdot m} \sum_{i=1}^m \sum_{k=1}^{n-1} (I_k^{\text{data}}(\bv{r}_i) - I_k(\bv{r}_i) )^2 + \underbrace{\frac{1}{n-1}\sum_{k=1}^{n-1} \lVert L \bv{c}_k \lVert_2^2}_{\text{regularization term}},
\label{eqn:mse_reg}
\end{equation}
for some suitably chosen (Tikhonov) linear transformation $L$ (\cite{tih99}), where $\lVert \cdot \lVert_2$ denotes $\ell^2$ norm. In this study we use 
\begin{equation}
L\bv{c}_k = \sqrt{\lambda}\beta\left( W \bv{c}_k - \bv{c}_k\right),
\label{eqn:tichonov}
\end{equation}
where $W$ is the interaction matrix (see Equation (\ref{eqn: exposure_mat})) and $\lambda>0$. 
%The purpose of regularization is to reduce model complexity by minimizing the differences in behaviour between nearby UTLAs (within the interaction region, recall that $\sigma$ = 10 km). 
%We explore the effect of the tuning parameter $\lambda>0$ in appendix \ref{app:reg}. 
We write the contribution to the mean squared error from the interval ending at $t=10k$, as
\begin{equation}
E_k = \frac{1}{m} \sum_{i=1}^m (I_k^{\text{data}}(\bv{r}_i) - I_k(\bv{r}_i) )^2 +  \lVert L\bv{c}_k \lVert_2^2
\label{eqn:singleerror}
\end{equation}
so
\begin{equation}
MSE_{\text{reg}} = \frac{1}{n-1}\sum_{k=1}^{n-1} E_k. 
\end{equation}
The critical ratios are determined by applying discrete gradient descent with momentum \cite{rum86} (see Equation (\ref{eqn:mom})) to each interval in turn. One may view critical ratios as biases that are adjusted as we fit the model to data. The fitting procedure is analogous to optimizing biases while training a neural network via gradient descent methods (\cite[5.2.1 Parameter optimization]{bish06}). 
When $E_k$ is sufficiently low, the descent algorithm is applied to the next interval. The descent method, with momentum term $v_k$, is
\begin{align} \label{eqn:mom}
\nonumber v_k(\bv{r})^{(0)} =& 0, \\ 
v_k(\bv{r})^{(j+1)} =& \mu v_k(\bv{r})^{(j)} + (1 - \mu) \nabla_k(\bv{r})^{(j)}, \\ \nonumber
c_k(\bv{r})^{(j+1)} =& c_k(\bv{r})^{(j)}
- \alpha v(\bv{r}, t)^{(j)},
\end{align}
where $\nabla_k(\bv{r}):= I_k(\bv{r}) - I_k^{\text{data}}(\bv{r})$ is a discrete gradient, $j$ indicates an iteration number (or an epoch), $\alpha >0$ is a learning rate, and $\mu>0$ is the exponential momentum decay rate.

We explore different choices of regularization parameter, decision temperature and the range of values of $c_k(\bv{r})$ parameters in \ref{app:reg}. The decision temperature cannot be measured directly, however we find that for $\beta>4$ the fitting method becomes unstable and errors become large as the model becomes \textit{subcritical}, meaning that two behavioural fixed points emerge (see section \ref{ssec:behdyn}). For very small $\beta$ values, corresponding to very large variations in behavioural parameters, we find higher $MSE$ and lower sensitivity to regularization. We select $\beta=0.5$ because this value maximizes sensitivity of spatial variation to regularization (reducing model complexity) and minimizes sensitivity of the $MSE$ to regularization. The regularization parameter is chosen using the ``knee'' method \cite{sat11} to be $\lambda=2$, in order to balance model complexity against model accuracy.

\subsection{Simulation Results}
\label{sec:sim}

\begin{figure}[h!]
	\centering
	\begin{tabular}{cc}
	(a) & (b) \\
		\includegraphics[scale=0.25]{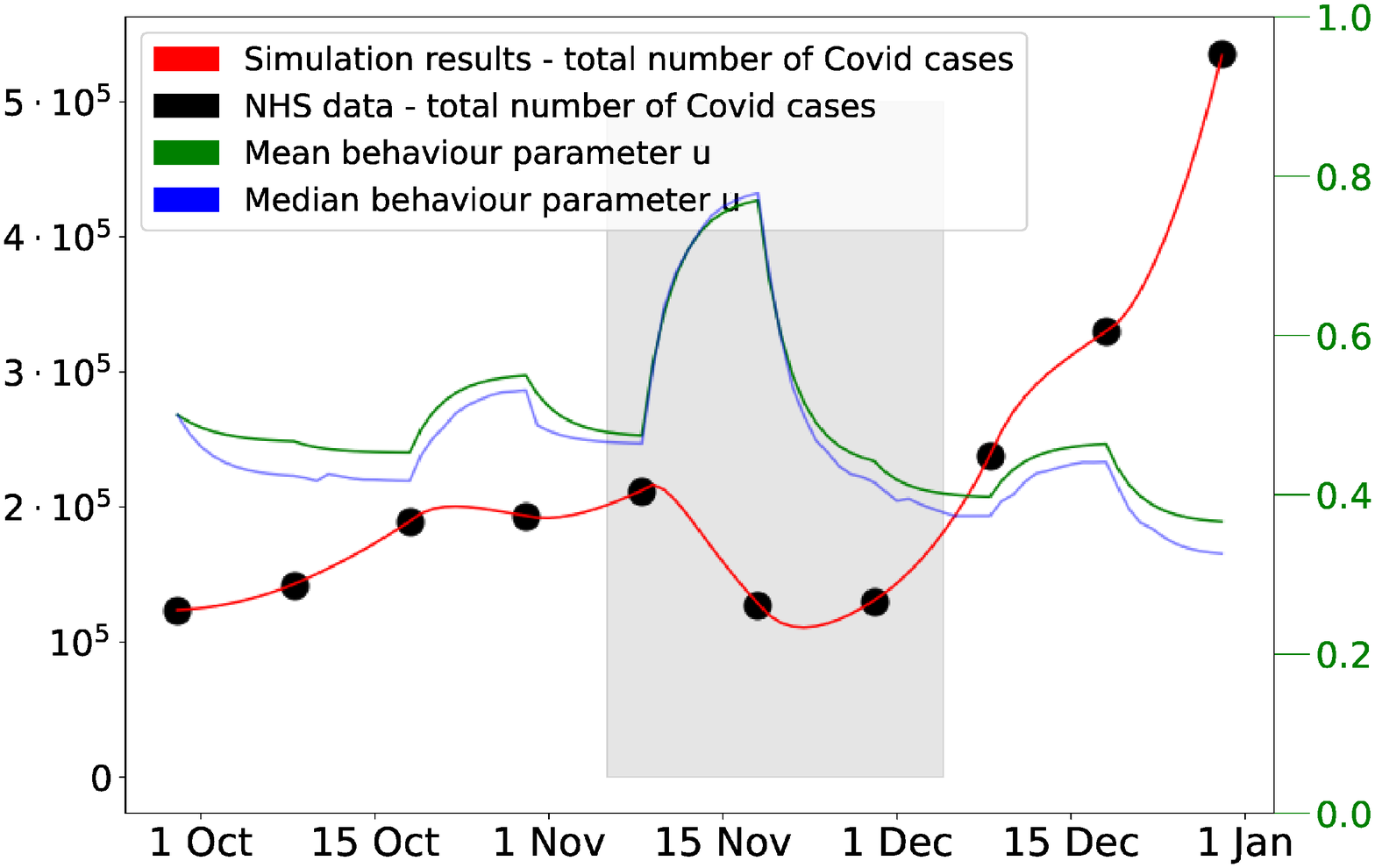} & 
	\includegraphics[scale=0.25]{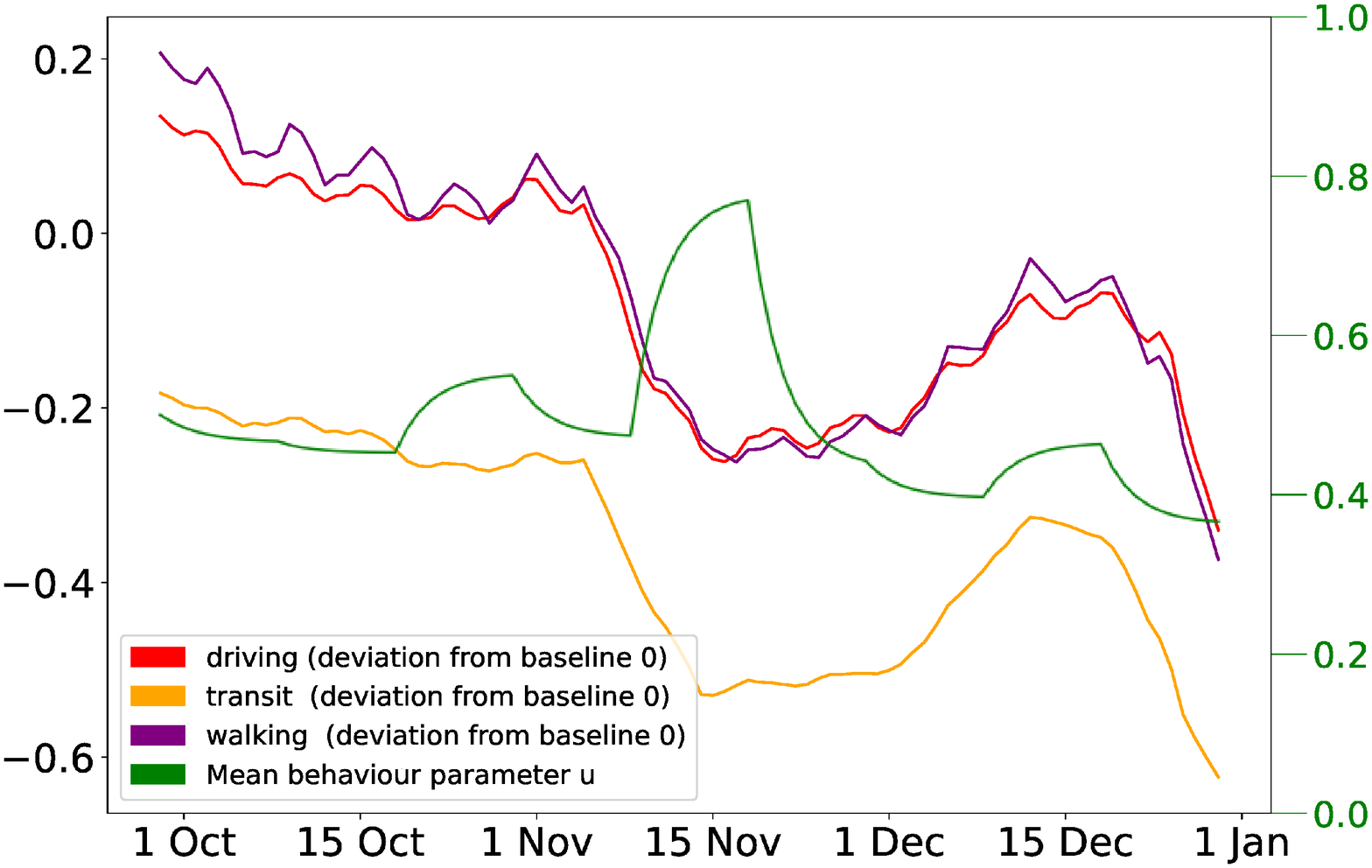}  \\
	\end{tabular} 
	\caption{(a) English active case numbers (black dots) October 2020 to January 2021. Grey band indicates national lockdown 5 November to 2 December 2020. Red and green curves indicate simulated infections and average implied cooperation rate, $u$ (equation (\ref{eqn:uav})), respectively.  (b) Levels of driving, transit and walking activity during the same period (Apple mobility reports \cite{app21}), compared to implied  cooperation rate. \label{fig:overallinf} }
\end{figure}

Figure \ref{fig:overallinf} (a) shows the active Covid-19 infection numbers \cite{covid_data}, with infection numbers from our fitted model (with $\beta=0.5$), and system average cooperation rate
\begin{equation}
u(t) = \frac{1}{m} \sum_{i=1}^m u(\bv{r}_i,t).
\label{eqn:uav}
\end{equation}
When the  cooperation rate is determined by calibration to disease data we refer to it as the \textit{implied cooperation  rate}. From Figure \ref{fig:overallinf} (a) we see that the calibrated model accurately reproduces infection rates, and during the lockdown period (marked by a grey zone), the implied  cooperation rate, $u$, jumps. However, there is a delay of approximately three days before this jump occurs. One explanation is that the onset of lockdown triggers a burst of activity by people wishing to relocate, make last minute visits to friends or relatives or make preparations for a period of confinement (such as shopping). People may also be unable to immediately start working from home, so do it gradually. This view is supported by the working from home in London graph in \cite{mob20}, and  Figure \ref{fig:overallinf} (b), where we compare the implied  cooperation rate  to Apple mobility reports \cite{app21} (driving, public transport and walking). Here we see that mobility falls gradually over 10-15 days following the onset of lockdown. When interpreting Figure \ref{fig:overallinf} we must also consider the possibility that our SIRS model doesn't adequately capture the incubation period (4-6 days \cite{McA20}), so in our model behavioural changes may influence infection rates faster than they do in reality, meaning that the implied behavioural change occurs later than in reality. Finally, Figure \ref{fig:overallinf} shows that cooperation appears to decline, and infections rise, before lockdown ends. This may represent lockdown fatigue or behavioural changes induced by reduced daylight and the buildup to Christmas, but it may also signal an increase in the underlying transmission rate of the virus due to the emergence of a new strain. We return to this point below.  

\subsubsection{Behavioural variations by location and wealth}
\label{sec:behwealth}

\begin{figure}[h!]
	\centering
	\begin{tabular}{cc}
\includegraphics[scale=0.175]{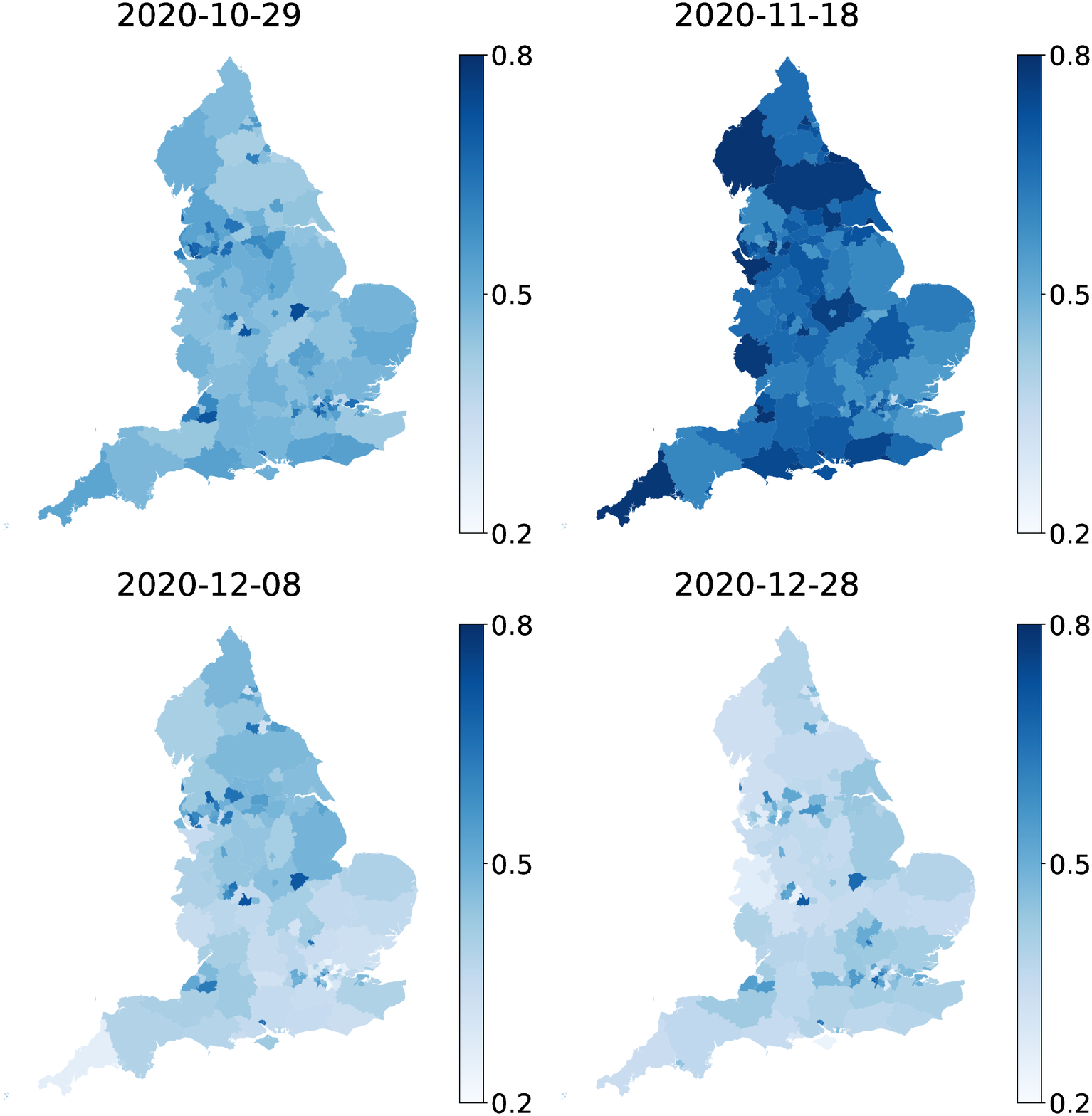} &
\includegraphics[scale=0.175]{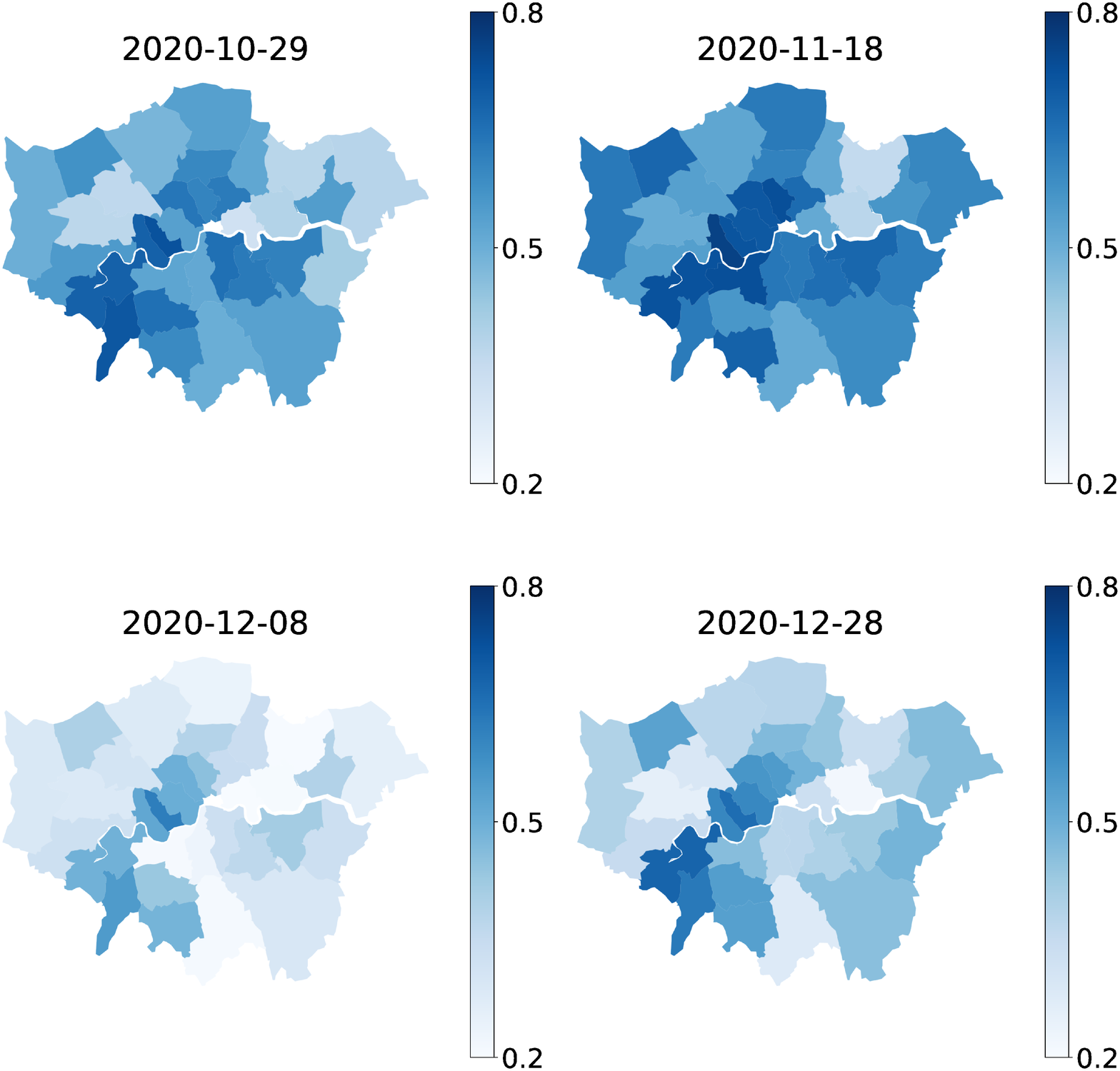} \\
(a) & (b) \\
	\end{tabular} 
	\caption{Implied  cooperation rate $u(\bv{r},t)$ during the period of interest in (a) England and (b) London. Dates were selected to illustrate cooperation behaviour before, during and after the second national lockdown in England. The date 2020-11-18 (18 Nov) lies in the middle of the lockdown interval (5 Nov 2020 to 2 Dec 2020) and we see that the corresponding map shows the highest levels of cooperation when compared to maps from outside the lockdown period. }
	\label{fig:umap}
\end{figure}

\begin{figure}[h!]
	\centering
\includegraphics[scale=0.4]{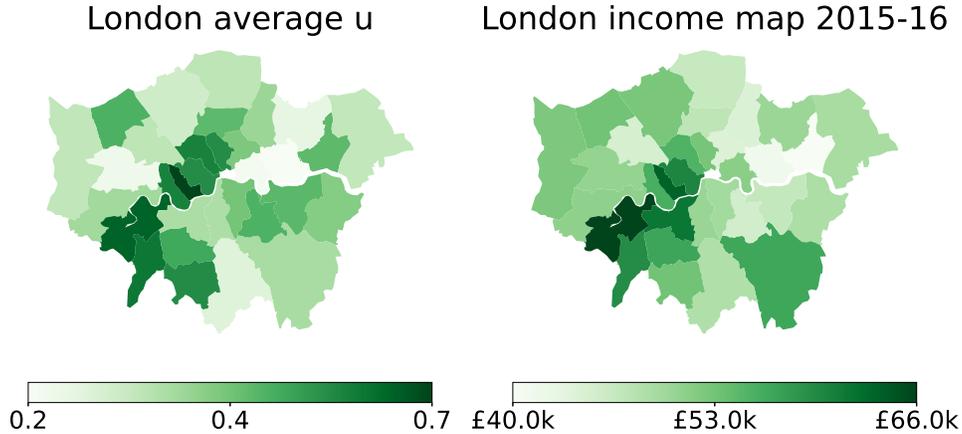} 
	\caption{Left map shows time-averaged implied cooperation $\langle u(\bv{r},t) \rangle_t$ in each London UTLA over the final three months of 2020. Right map shows mean household income in the same UTLAs in 2015-2016. The City of London, UTLA code E09000001, located in the centre of both maps, was excluded from the map due to insufficient active cases data. }
	\label{fig:wmap}
\end{figure}

\begin{figure}[h!]
	\centering
\includegraphics[scale=0.4]{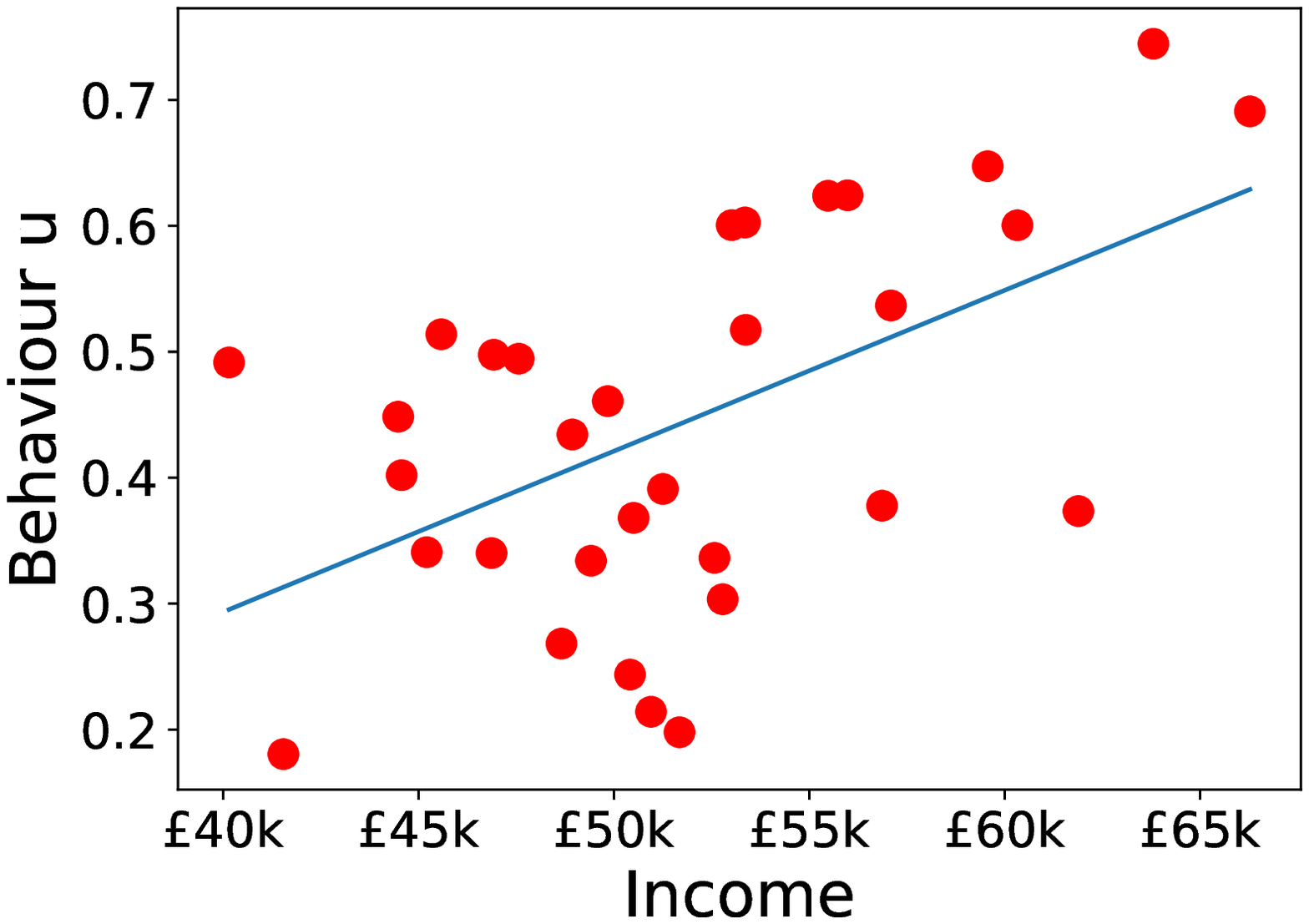} 
	\caption{Scatter plot of time-averaged implied  cooperation  $\langle u(\bv{r},t) \rangle_t$ in each London UTLA over the final three months of 2020 against mean household income during 2015-2016. Blue line shows linear regression model (\ref{eqn:regression}) fitted to data, having intercept $\hat{\alpha}_0 = -0.217$ and gradient $\hat{\alpha}_1=0.0128$.   }
	\label{fig:uvsw}
\end{figure}

Figure \ref{fig:umap} shows spatio-temporal variations in implied  cooperation rate in (a) England and (b) London. Figure \ref{fig:umap} (a) shows that during lockdown south east England exhibited lower implied  cooperation rates, while very rural UTLAs (such as Cornwall and the Lake District) have higher rates. However, the fact that sparsely populated areas naturally have higher levels of social distancing may partly explain this. Figure \ref{fig:umap} (b) shows how implied cooperation rates vary in London during the study period. Readers familiar with this city will notice that the traditionally wealthier southwest and central areas appear to exhibit higher cooperation levels. To quantify this effect we introduce the time averaged implied cooperation rate in cell $\bv{r}$
\begin{equation}
    \la u(\bv{r},t) \ra_t = \frac{1}{n} \sum_{k=0}^{n-1} u(\bv{r},10k).
\end{equation}
The values of this time averaged rate for London UTLAs are shown in Figure \ref{fig:wmap}, along side mean income levels in the same areas from 2015-2016 \cite{wea18}. While the visual similarity is clear, we can also quantify the relationship between wealth and implied  cooperation by fitting a linear model of the form
\begin{equation}
    \la u(\bv{r},t) \ra_t = \alpha_0 + \alpha_1 w(\bv{r}) + \varepsilon
    \label{eqn:regression}
\end{equation}
where $w(\bv{r})$ is mean annual income per person in thousands of pounds in cell $\bv{r}$ and $\varepsilon$ is normal error. The raw data and fitted linear relationship are shown in Figure \ref{fig:uvsw}, where coefficient estimates are $\hat{\alpha}_0 = -0.217$ and $\hat{\alpha}_1=0.0128$, with $R^2=0.283$, corresponding to a correlation between implied behaviour and income of $\mathrm{Cor}(\la u(\bv{r},t) \ra_t,w(\bv{r})) = 0.53$. The $t$-statistic for the gradient $t=\hat{\alpha}_1/\text{se}(\hat{\alpha}_1)=3.438$ and corresponding $p$-value $p=0.002$ provide strong evidence against the null-hypothesis of no relationship between income and implied cooperation. In other words, the relationship between implied behaviour (cooperation) and income is statistically significant. Higher implied rates in the wealthier areas are consistent with the observation that the relative cost of  cooperation for poorer members of society is higher, leading to lower levels of compliance \cite{ken21}. In addition, the nature of many less-well paid jobs makes  infection-reducing behaviours such as social distancing more difficult or sometimes impossible to carry out \cite{weill20}.

\subsubsection{Impact of the tier system on behaviour}
\label{sssec:tier}

\begin{figure}[h!]
	\centering
		\begin{tabular}{cc}
\includegraphics[scale=0.25]{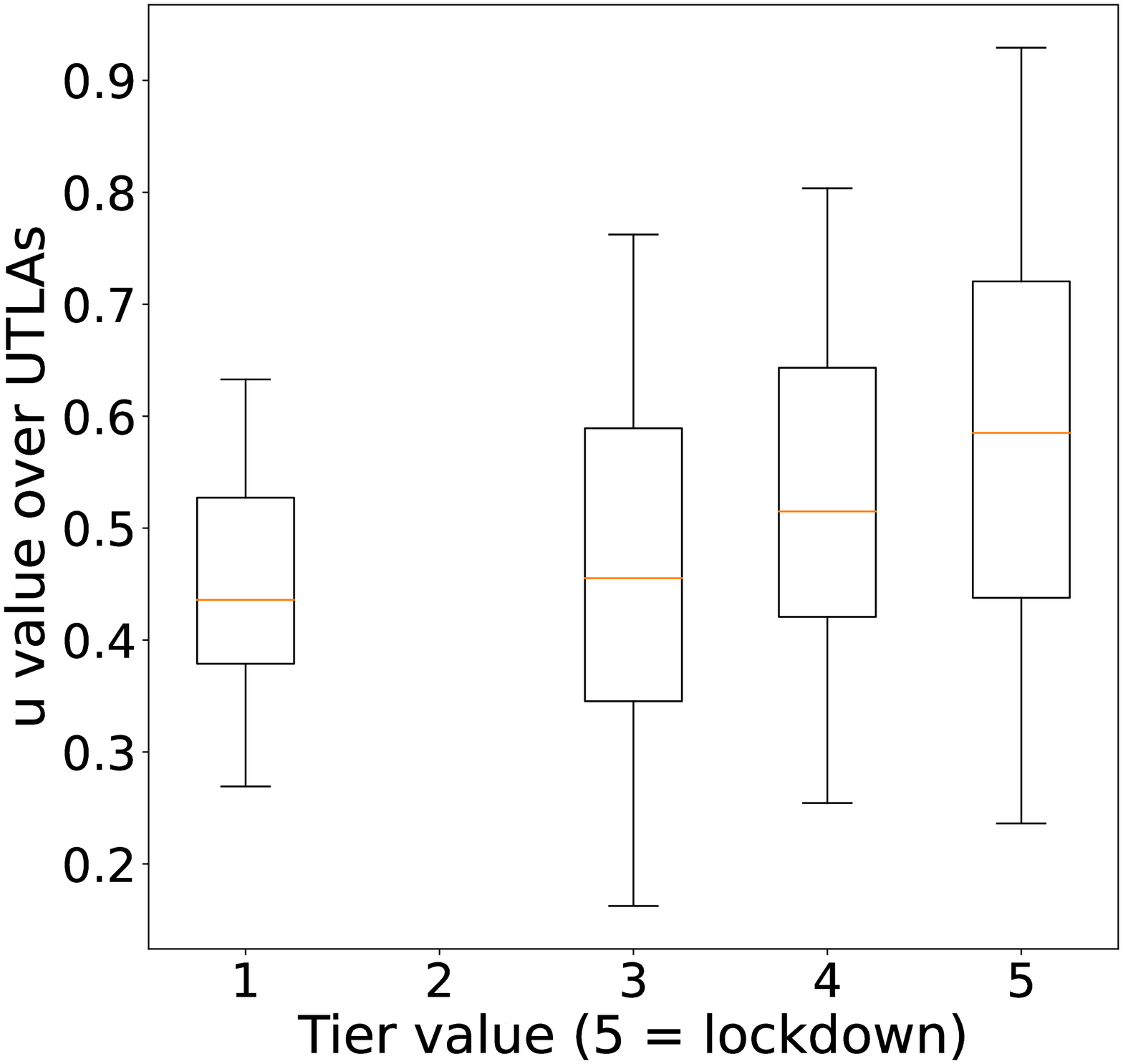} &
\includegraphics[scale=0.45]{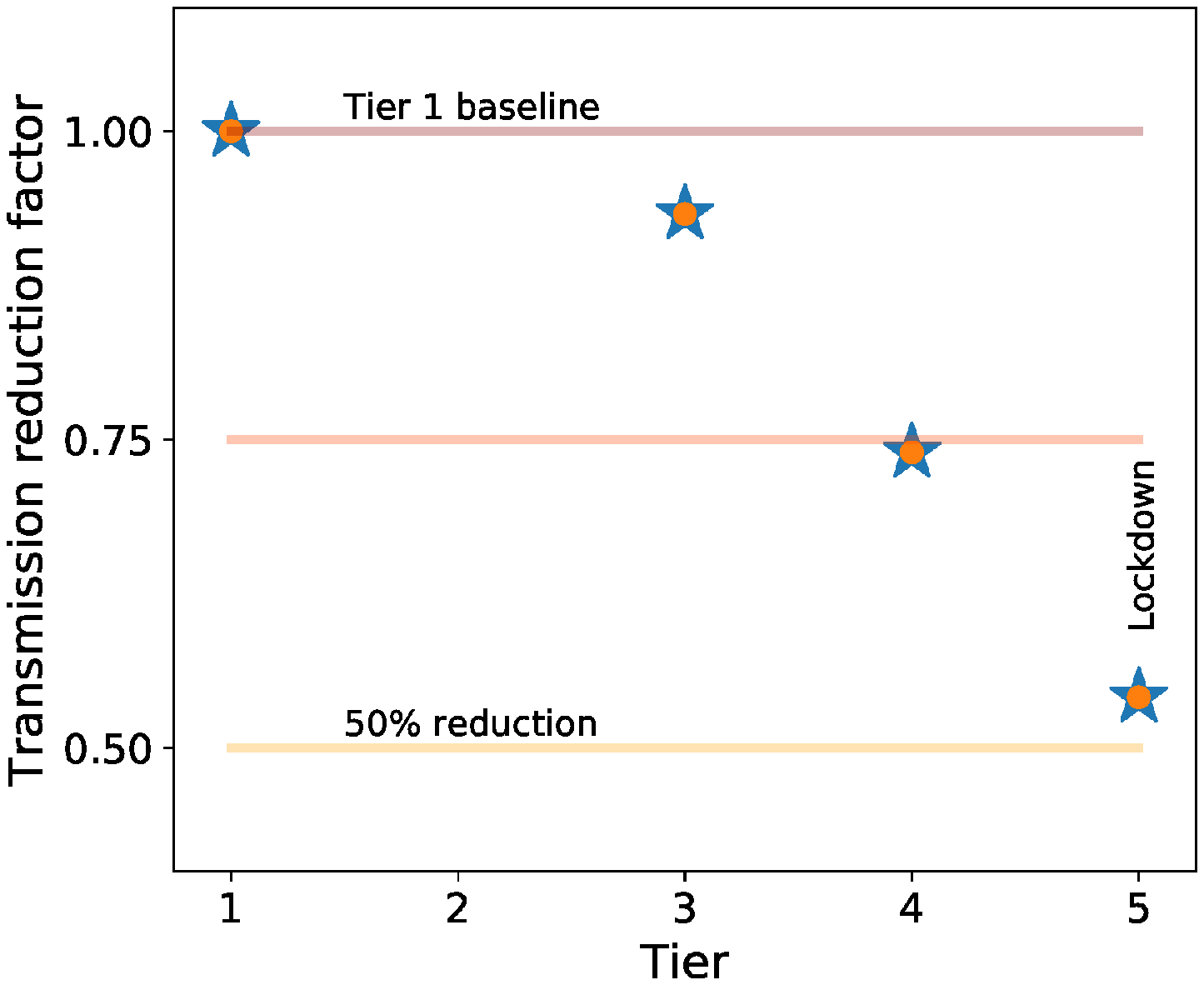} \\
(a) & (b) \\
	\end{tabular} 
	\caption{(a) Distribution of behaviour states in each lockdown tier between 12 October and 18 December 2020. The orange line represents the median, and the boxes extend from the lower to the upper quartile of the data. The whiskers that extend from the box determine the range of the data. There are no values in Tier 2 during the selected period. (b) Median transmission reduction factor compared to tier 1 baseline, $\rho_{\text{rel}}(T)$, in each tier, $T$. See equation (\ref{eqn:rho}) for definition. }
	\label{fig:tiers}
\end{figure}

A potential application of our model is to understand the impact of government restrictions on behaviour.  In October 2020, regional variations in these restrictions were formalized as a national tier system, ranging from tier 1 medium alert (maximum of six people meeting indoors, retail open, table service only in restaurants) up to full lockdown (which we assign to tier value 5, for the consistency with the other 4 tiers). Figure \ref{fig:tiers} shows how the distribution of implied  cooperation rate varies between tiers. While behaviour does respond to the highest level of restrictions, there is little difference between tiers 1 and 3, suggesting that these finer gradations had little material effect on  cooperation (or the disease). It seems that the public were either not aware of the differences between the restrictions in each tier, or largely ignored them, or that the differences had little effect on disease transmission. In order to quantify how behavioural changes induced by the tiers affect disease transmission, we define the within-cell transmission reduction factor
\begin{equation}
    \rho(\bv{r},t) = (1-u(\bv{r},t))^2
\end{equation}
which gives the factor by which infection rates \textit{between} members of cell $\bv{r}$ are reduced by cooperation at time $t$. Although the geographical spread of the disease depends on cooperation rates in \textit{neighbouring} cells, this within-cell metric provides a simple way to understand the effect of the tier system on disease transmission rates. We write the average of the median transmission reduction factors over all cells in tier $T$ over our period of study as $\bar{\rho}_T$, then the reduction factor relative to the tier 1 baseline is
\begin{equation}
    \rho_{\text{rel}}(T) = \frac{\bar{\rho}_T}{\bar{\rho}_1}.
    \label{eqn:rho}
\end{equation}
The values of this quantity for each tier are displayed in Figure \ref{fig:tiers} (b), where we see that while there is little difference in transmission rates between tiers 1 and 3, tiers 4 and 5 have a substantial impact, with tier 5 median disease transmission rates reduced by $50\%$ with respect to tier 1.

\subsubsection{Disease's geographical progression}

\begin{figure}[h!]
	\centering
	\begin{tabular}{l}
		(a)
		\\
		\includegraphics[scale=0.3]{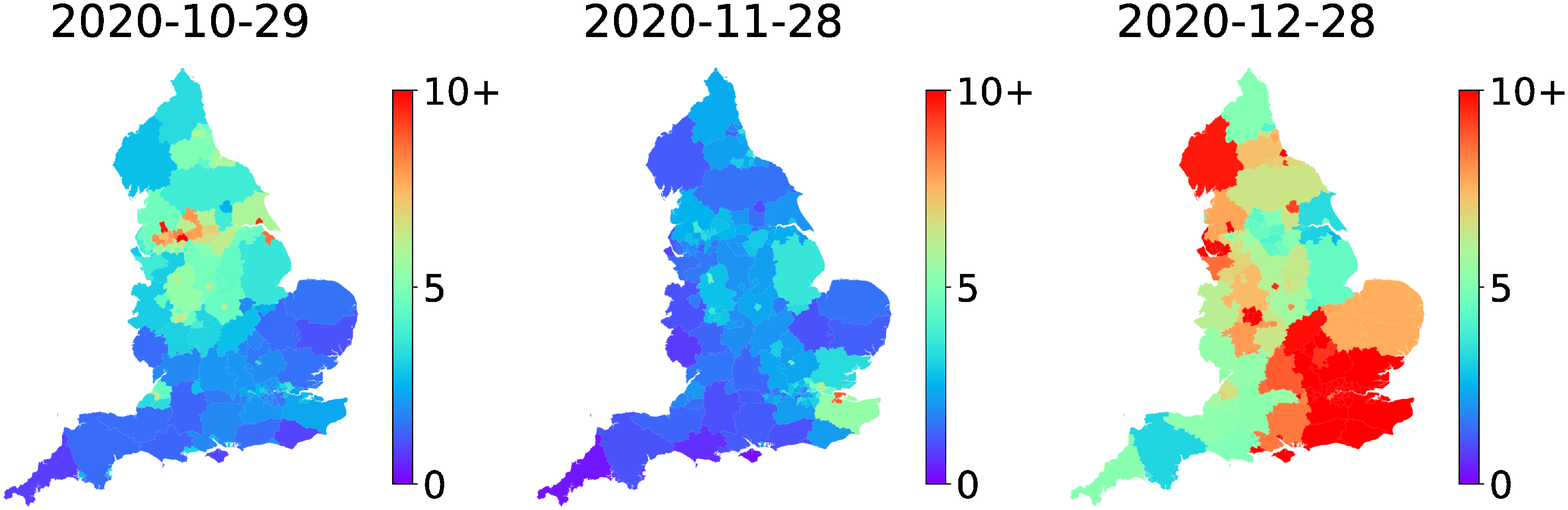}\\
		(b)
		\\
		\includegraphics[scale=0.3]{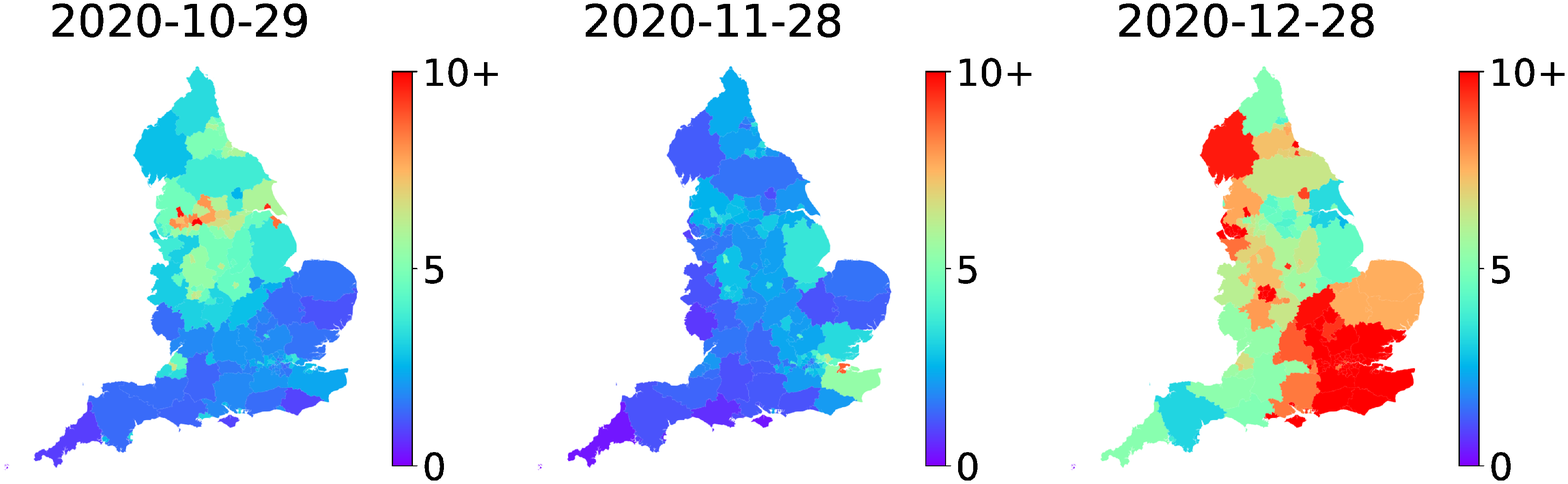}
	\end{tabular}
	\caption{(a) The choropleth map of England (split into UTLAs) presenting the active Covid-19 cases per 1000 people in England on 29 October, 28 November and 28 December 2020. (b) Active cases obtained from simulations of our model using the calibrated critical ratios $c_k(\bv{r})$ with $\beta=0.5$, $\lambda = 2.0$. The differences between these two maps are captured by Figure \ref{fig:diffmaps} in \ref{app:reg}.}
	\label{fig:maps}
\end{figure}

We now consider how well our model captures the disease's geographical progression.  To avoid  maps being biased by population densities we show the number of active cases per 1000 people in each UTLA. Figure \ref{fig:maps} (a) shows spatio-temporal variations in case numbers from NHS data \cite{covid_data}. Figure \ref{fig:maps} (b) shows simulated case numbers. The two maps are visually indistinguishable and we find that the absolute difference in cases between data and simulation is, on average, $\approx 3$ cases per cell (cf. average cell population of $3.7 \times 10^5$). See Figure \ref{fig:diffmaps} in appendix \ref{app:reg} for spatial maps of these differences.  Figure \ref{fig:maps} also shows a growth in the number of cases in Kent (compared with other regions) in early December 2020. This is explained by Lineage B.1.1.7 (\cite{kent20, kir20}) which caused a spike of infections in December in England due to its higher transmissibility \cite{voltz21}. 

\subsubsection{Summary and value of approach}

We are able to accurately capture high resolution data describing the spread of infection through a spatial domain (England) by calibrating a simple spatial game model of  infection-reducing (cooperative) behaviour, in which transmission of disease between cells depends on the behavioural states in those cells, as well as their proximity. We have presented two applications of our approach, showing that  infection-reducing behaviour (or the ability to behave in this way) is significantly correlated to income. We have also shown how the tier system impacts behaviour and transmission rates. The fact that spatial variation in transmission rates, and the effects of intervention, are partially predictable, raises the importance of spatial behaviour modelling. For example, our model would in principle allow us to discover though numerical experiments which cells contribute most significantly to the spread of disease, leading to spatially targeted restrictions which exploit known demographic or geographical factors in order to create low transmission rate regions which act as \textit{barriers} to spread. In addition, regression of our calibrated critical ratios against proxy behaviour measures such as mobility and social media data offer the possibility of incorporating such data into disease modelling, and prediction.

As with most models of social systems, major simplifications have been made. Our implied  cooperation rates should be viewed in a similar way to implied parameters in other social or economic models. For example, in mathematical finance, implied volatilities \cite{impvol1,impvol2} are widely used as a guide to asset behaviour, but do not precisely capture the true volatility or price processes. The value of modelling approaches such as ours, is the insights they provide into how population level behaviour varies spatially, is affected by government advice or rules, by demographic factors, and by infection rates (section \ref{sec:delay}).

\section{Instability at critical information delay}
\label{sec:delay}

\begin{figure}%[tbhp]
	\centering
	\begin{tabular}{cc}
	(a) & (b) \\
		\includegraphics[width=0.45\linewidth]{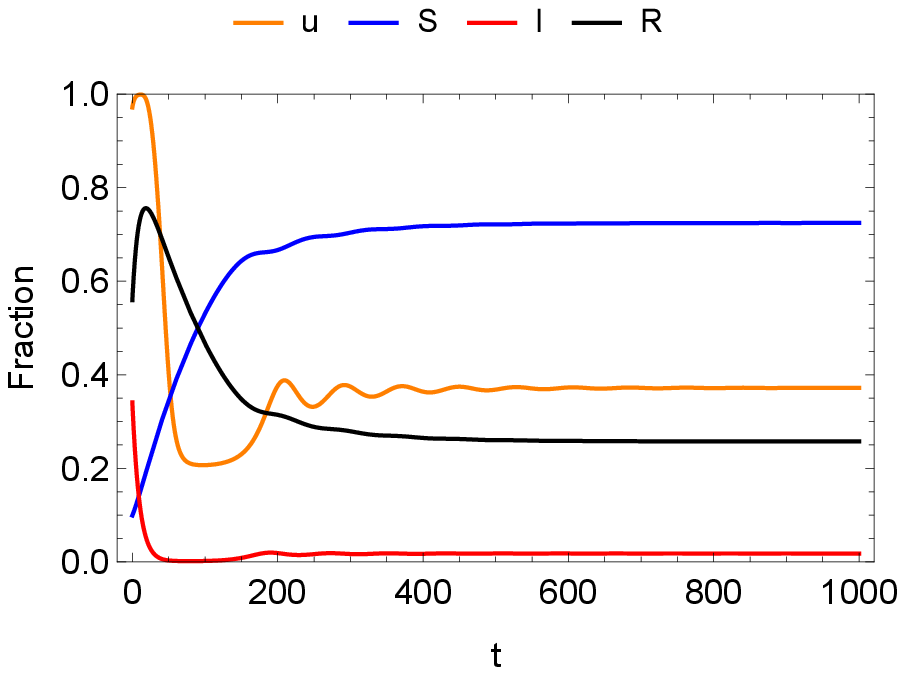}
		&
		\includegraphics[width=0.45\linewidth]{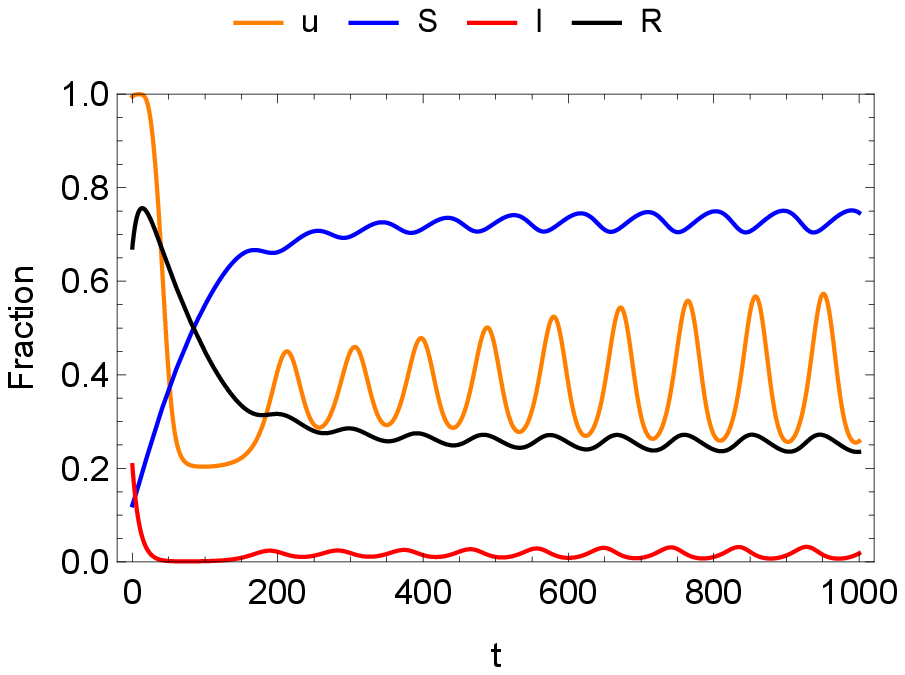}
	\end{tabular}
%	\centering
%	\includegraphics[width=0.5\linewidth]{BSIRS_stable_T_10_beta1_3_c1_20.eps}
	\caption{Time evolution of the four variables $u$ (cooperation rate),  $S$ (susceptible fraction),  $I$ (infected fraction), $R$ (recovered fraction) of the BSIRS process. Fixed parameters of the behaviour model are (a) $c_0=2,c_1=20,\beta_1=3$ and delay $T=10$; (b) $c_0=2,c_1=20,\beta_1=3$ and delay $T=15$. All other parameters are as in Table \ref{tab:par}. In plot (a), the delay time $T=10$ is below the critical threshold $T_c=12.9$ for the given parameter values, so initial oscillations vanish, leading to a stable steady state of behaviour and disease variables. In plot (b) the delay time, $T=15$, is above the critical threshold and the fixed point of the dynamics is destabilized, leading to oscillations in behaviour and disease parameters.} 
	\label{fig:BSIRS_sols}
\end{figure}

Previously we considered how behaviour affected disease spread. We now consider theoretically the possibility that disease can affect behaviour in return. We define a non-spatial version of the BSIRS model, allowing behavioural parameters to depend on recent rates of infection. Formally we set $W(\bv{r},\bv{r}')=\delta_{\bv{r},\bv{r}'}$, where $\delta_{\bv{r},\bv{r}'}$ is the Kronecker delta, and consider our system to consist of a single cell, allowing us to drop $\bv{r}$ dependence from  equations (\ref{eqn:du}) to (\ref{eqn:dR}), yielding
\begin{align}
\label{eqn:del1}
\dot{u} &= \frac{1}{\tau}(F(u) - u) \\
\dot{S} &= \xi R- \epsilon (1-u)^2I S  \\
\dot{I} &= \epsilon (1-u)^2I  S - p I \\
\dot{R} &= p I - \xi R.
\label{eqn:del2}
\end{align}
For analytical convenience we use following parameterization of the response function
\begin{equation}
F(u) = \ept \left(-\beta_0 + \beta_1 u \right) 
\end{equation}
where $\ept = \lgt^{-1}$. As in our simulations we assume that the coefficient, $\beta_1$, of $u$ is constant, but we allow the scaled critical ratio $\beta_0 = \beta c$ to vary with time. It is time variations in this parameter due, for example, to government restrictions or raised risk levels, which drive behavioural change. Assuming that behaviour, $u(t)$, is at a fixed point $u^\ast$, we have the following fixed point for the disease variables
\begin{align}
S^\ast &= \frac{p}{\epsilon(1-u^\ast)^2 }, \\ 
I^\ast &= \frac{ (\epsilon(1-u^\ast)^2 -p)\xi}{\epsilon(1-u^\ast)^2  (p + \xi)}, \\
R^\ast &= \frac{ (\epsilon(1-u^\ast)^2 -p)p}{\epsilon(1-u^\ast)^2  (p + \xi)}.
\end{align}
We now examine the stability of this fixed point when $\beta_0$ is reduced by higher infection rates. Since information about current risk levels may be delayed due a time lag in the reporting of national case rates, or the public not keeping up to date with government announcements, we assume the following relationship
\begin{equation}
\beta_0 = c_0 - c_1 I(t-T)
\end{equation}
where $c_0$ is the scaled critical ratio in the absence of infection, $c_1>0$ measures sensitivity to infection rates, and $T>0$ is a delay time. The response model is now treated as a function of two variables
\begin{equation}
F(u,I) = \ept \left(-c_0 + c_1 I + \beta_1 u \right) 
\label{eqn:puI}
\end{equation}
where time dependence is suppressed for brevity. Equation (\ref{eqn:puI}) is the most general form of a logistic linear model for response in terms of infection rates and behaviour. If we were to allow infection dependence in $\beta_1$, this would introduce cross terms between behaviour and infection, and we would no longer have a generalized linear model. On the basis that the simplest explanations of phenomena should be considered first (Occam's razor \cite{mac03}) we analyse the linear model. In many game models, delaying the time at which players receive payoff information leads to instability in the evolution of population strategies \cite{mie11,bur17,yi97}. We can predict the onset of this instability \cite{ern09} by considering perturbations of the BSIRS variables about a fixed point
\begin{equation}
\begin{pmatrix}
u \\
S \\
I \\
R 
\end{pmatrix}
=
\begin{pmatrix}
u^\ast \\
S^\ast \\
I^\ast \\
R^\ast 
\end{pmatrix}
+
\begin{pmatrix}
\delta u \\
\delta S \\
\delta I \\
\delta R 
\end{pmatrix}.
\end{equation}
We define $\bv{x} = (u,S,I,R)^T$ so our perturbations may be written $\bv{x}=\bv{x}^\ast + \delta \bv{x}$. The linearised equations for these perturbations take the form $\delta \bv{x}(t) = M \delta \bv{x}(t) + B \delta \bv{x}(t-T)$ where $M$ and $B$ are constant matrices. Denoting the $u$ and $I$ derivatives of the response function by $F_u$ and $F_I$, we have
\begin{equation}
M =\begin{bmatrix}
\frac{1}{\tau}(F_u(u^\ast,I^\ast)-1) & 0 & 0 & 0 \\
2\epsilon (1-u^\ast)I^\ast S^\ast & -\epsilon (1-u^\ast)^2 I^\ast & -\epsilon (1-u^\ast)^2 S^\ast & \xi\\
-2\epsilon (1-u^\ast)I^\ast S^\ast & \epsilon (1-u^\ast)^2 I^\ast & \epsilon (1-u^\ast)^2 S^\ast-p & 0 \\
0 & 0 & p & -\xi
\end{bmatrix}
\end{equation}
and
\begin{equation}
B  = \begin{bmatrix}
0 & 0 & \frac{1}{\tau}F_I(u^\ast, I^\ast) & 0 \\
0 & 0 & 0 & 0 \\
0 & 0 & 0 & 0 \\
0 & 0 & 0 & 0 \\
\end{bmatrix}.
\end{equation}
Denote by $\mathbf{1}$ the $4 \times 4$ identity matrix. We now seek a solution to our linearised equation of the form
\begin{equation}
\delta \bv{x}(t) = e^{z t} \bv{v}
\end{equation}
where $z \in \mathbb{C}$ and $\bv{v} \in \mathbb{C}^4$ is a constant complex vector. Noting that $\delta \bv{x}(t-T) = e^{-z T} \delta \bv{x}(t) $, and $\dot{\delta \bv{x}}= z \delta \bv{x}$ we have $\left(z \mathbf{1} - M - B e^{-z T} \right) \delta \bv{x} = 0$, so
\begin{equation}
\det \left(z \mathbf{1} - M - B e^{-z T} \right)=0,
\label{eqn:det}
\end{equation}
which is the \textit{characteristic equation}. For a given set of behaviour and disease parameters, the solution, $z^\ast(T) \in \mathbb{C}$, to the characteristic equation may be viewed as a function of the delay, $T$. If $z^\ast(T)$ has negative real part, then the fixed point is stable, whereas positive real part implies instability. An example of the transition from stability to instability as $T$ increases beyond a critical threshold $T_c$ is shown in Figures \ref{fig:BSIRS_sols} (a) and \ref{fig:BSIRS_sols} (b).
The appearance of periodic solutions surrounding an equilibrium point as a system parameter varies is known as a \textit{Hopf Bifurcation} \cite{mar76,ern09}. We can find the bifurcation point numerically by noting that the real part of $z^\ast(T)$ will be zero at the transition. We therefore set $z = i \omega$ (zero real part) in equation (\ref{eqn:det}), and separate into real and imaginary parts, yielding two simultaneous equations in the two real parameters $\omega$ and $T$ (assuming all other parameters fixed). The critical parameter values $T_c$ and $\omega_c$ which solve these simultaneous equations are those for which the characteristic equation admits a solution with zero real part. The value $T_c$ is the critical delay time at which oscillations emerge ($\omega_c$ is their angular frequency). 
\begin{figure}[h]%[tbhp]
	\centering
	\includegraphics[width=0.5\linewidth]{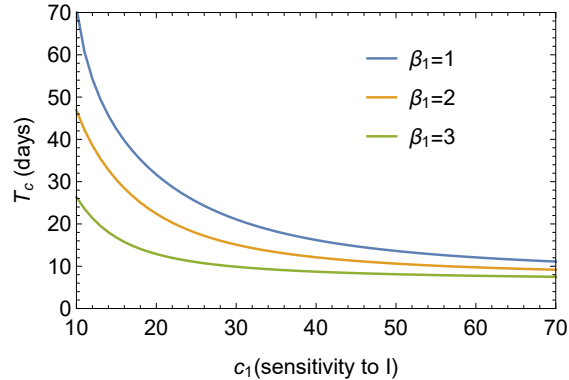}
	\caption{Relationship between critical delay $T_c$ (the minimal length of the delay from which the oscillations for $u$, $S$, $I$ and $R$ occur, see Figure \ref{fig:BSIRS_sols}) and sensitivity, $c_1$, of behaviour to infection rates for BSIRS process with $c_0=2,\beta_1 \in \{1,2,3\}$ and other parameters given in Table \ref{tab:par}. }
	\label{fig:Tc}
\end{figure}
In Figure \ref{fig:Tc} we have used this method to study how the critical delay depends on  behavioural parameters $c_0,c_1,\beta_1$, with disease transmission parameters given by Table \ref{tab:par}. We fix $c_0=2$, yielding a  cooperation rate in the absence of infection $u \approx 12\%$. That is we assume that some basic level of  infection-reducing behaviour is the norm (this is not a new phenomenon; the slogan ``coughs and sneezes spread diseases'' appeared in the 1918 influenza pandemic and has remained in use). The sensitivities $c_1, \beta_1$ to case rates and the behaviour of others are unknown, and we therefore explore a range of values: $c_1 \in [10,70]$ and $\beta_1 \in \{1,2,3\}$. We have verified that our critical delay values (Figure \ref{fig:Tc}) are correct by direct solution of the delay equations (\ref{eqn:del1})-(\ref{eqn:del2}) above and below the predicted thresholds $T_c$. For example, for the parameter values used in Figure \ref{fig:BSIRS_sols} (a) and (b) the critical delay point is $T_c=12.9$ days. Figure \ref{fig:BSIRS_sols} (a) shows a stable case $T=10$ and in Figure \ref{fig:BSIRS_sols} (b) the delay time $T=15$ exceeds the bifurcation point, so behavioural oscillations appear leading to recurring peaks and troughs of infection. 

From Figure \ref{fig:Tc} we see that provided behavioural sensitivity to infections is low, then provided the public are not more than  $\approx 1$ month out of date in their perception of disease rates, then the disease and behaviour should reach equilibrium. As the sensitivity increases, delays greater than 10 days can create instability. This suggests that in the long term, if Covid-19 becomes an endemic disease controlled by voluntary cooperation (rather than government imposed measures) then it will be important to make the public aware of infection rates in a timely and clear fashion. However, this suggestion is based on our idealized mathematical model, and we acknowledge the public behaviour can be much more complex and unpredictable than our modelling assumptions imply.

\section{Conclusion}

The current pandemic has generated new interest in the interaction between disease and behaviour \cite{bau04,rel10,bha19,ara21,gou21,mwa20,gio20,sil19,mah20,vru20,giu20,tso21}. Analysis has ranged from the investigation of standard compartmental models where the transmission rate is allowed to depend deterministically on infection levels, to sophisticated spatial models which adapt techniques from modern physics to model human motion and interactions \cite{vru20,sil19}. While some models have been compared to non-spatial data, spatial analysis has tended to be theoretical rather than data-driven. Spatial modelling is important because infection rates and interventions vary between locations, and because the case numbers in one location will depend on past behaviour in those that surround it. Inferring how  infection-reducing behaviour has evolved therefore requires us to model how spatio-temporal behavioural variations have produced observed case numbers. In this work we have defined a minimal spatial game model, equivalent to the Hopfield neural network \cite{hop82,hop84,mac03}, which is coupled to the spatial SIRS model \cite{ker27}, and we have used our model to infer behavioural dynamics from high resolution spatio-temporal case data \cite{covid_data}. Our inferred dynamics can accurately reproduce the history of the Covid-19 pandemic toward the end of 2020. As well as modelling how behaviour affects disease, we have studied theoretically how behavioural responses to disease data can change system dynamics. By assuming that public access to information about case numbers may be delayed, we showed that there is a critical delay time beyond which behavioural dynamics destabilizes, leading to oscillations in case numbers.

Models in the social sciences which generate implied system parameters \cite{bur21,impvol1,impvol2} are useful for several reasons. In disease transmission, they provide a model-based quantitative link between non-measurable  infection-reducing (cooperative) behaviour, and measurable disease case rates. They can therefore provide a quantitative prediction of how people have been behaving, based on the outcomes of that behaviour. As simple applications, we have shown how implied infection-reducing behaviour (cooperation) was affected by the tier system, and how it varies across the capital city (richer areas cooperate more). Further, data sources such as mobility indices \cite{mob20}, retail data, mobile phone records \cite{mobR20}, social media output, crime rates and google reviews, can serve as proxy measurements for behaviour. In future work, regression of implied behavioural parameters on these proxy measurements might then allow us to quantitatively link behaviour data to disease spread. The ability to do this may be valuable, because proxy behavioural data is in plentiful supply and may be partially predictable, offering the possibility of advanced warning of future disease hot spots.

\section*{Data and Code availability}
All the data and simulation results can be found on the dedicated GitHub repository:
\newline \url{https://github.com/gnacikm/BSIRS_model}.

\section*{Acknowledgments}	
J.B. is  grateful for the Royal Society APEX
Award APX\textbackslash R1\textbackslash 180117 which supported this work.
The authors are grateful to Samia Burridge for her comments regarding the applications and readibility of the paper. The authors would like to thank to Riccardo Di Clemente for useful discussions regarding mobility reports during Covid-19 pandemic, and Cuebiq for supplying location data which guided our thinking in the early stages of this work.

\appendix

\section{Choice of regularization parameter and decision temperature} 

\label{app:reg}

\subsection*{Effects of regularization}

\begin{figure}[h!]
	\begin{tabular}{cc}
		\includegraphics[width=0.5\linewidth]{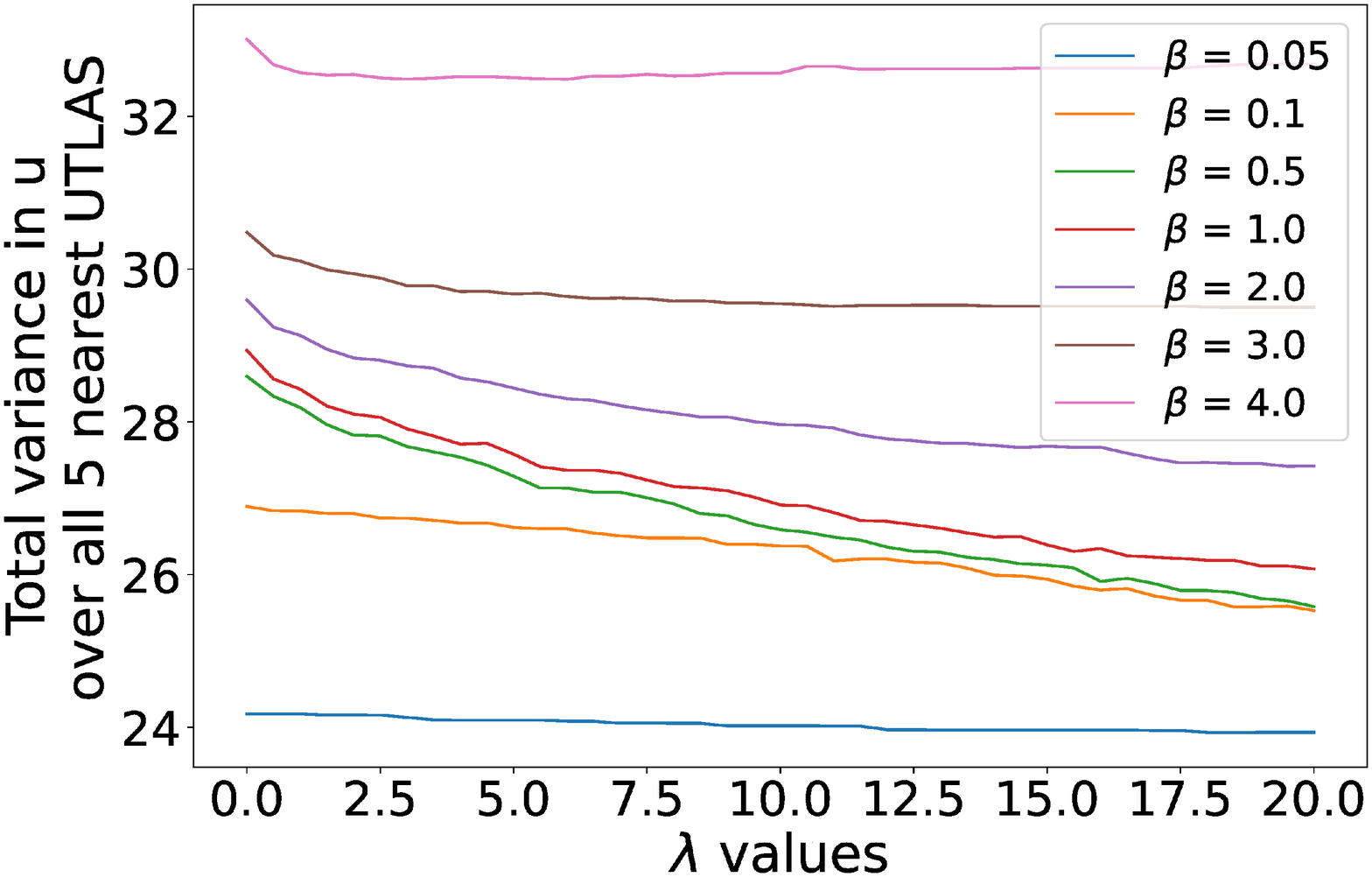} & \includegraphics[width=0.5\linewidth]{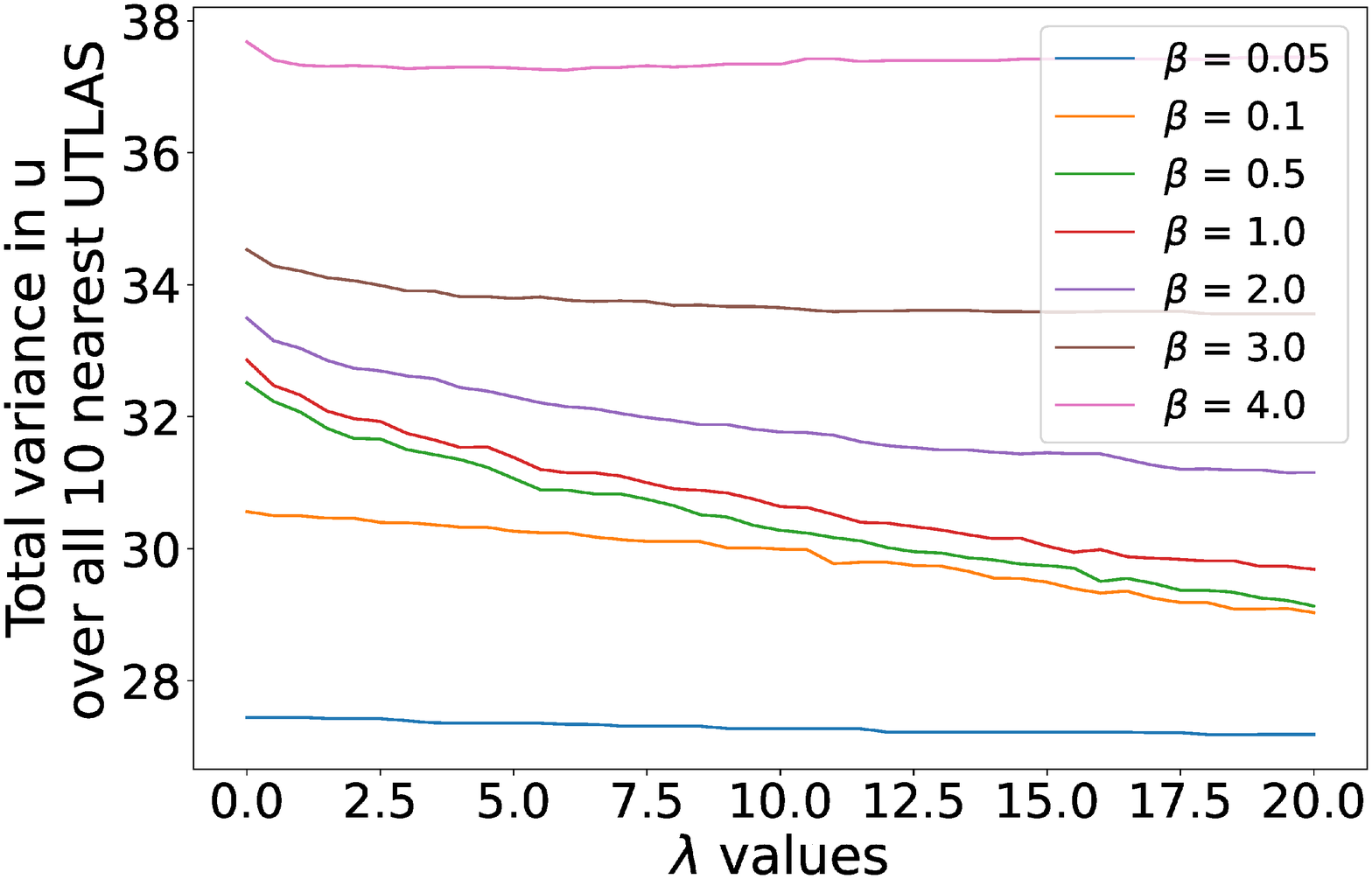} 
	\end{tabular}
	\caption{Effect of regularisation on similarity in behaviour (values of $u$) in the 5 (LHS) and 10 (RHS) nearest neighbouring UTLAs for different $\beta$ values. We can observe the decrease of the variance as $\lambda$ increases. The total variance is calculated as $\sum_{i=1}^{m} \sum_{k=1}^9 \mbox{Var}[u_{K}(\bv{r}_i, 10k)]$, where $u_{K}(\bv{r}, 10k)$ is a vector of $u$ values of $K \in \{5, 10\}$ nearest neighbours of $\bv{r}$ (obtained from the $W$ matrix) at time period $10k$. }
	\label{fig:reg_neigh}
\end{figure}

\begin{figure}[h!]
	\centering
	\includegraphics[width=0.7\linewidth]{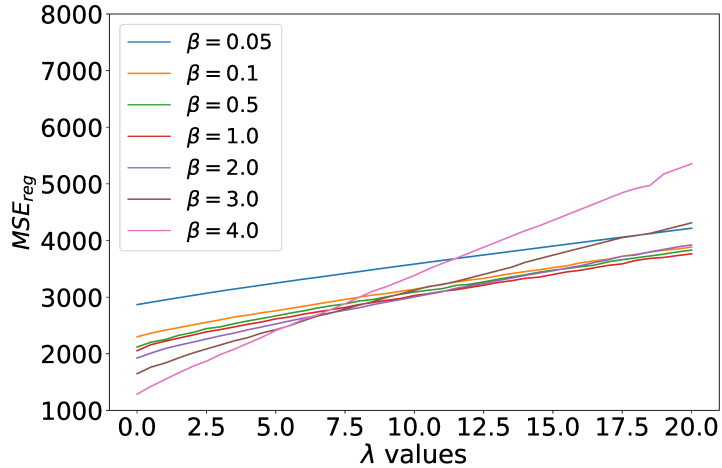} 
	\caption{Effect of regularisation on the fitting score $MSE_{\text{reg}}$. Note that about $\lambda=7$ all  $MSE_{\text{reg}}$ scores for different $\beta$ value are similar. }
	\label{fig:reg_scores_lam}
\end{figure}

\begin{figure}[h!]
	\begin{tabular}{lll}
		\includegraphics[width=0.31\linewidth]{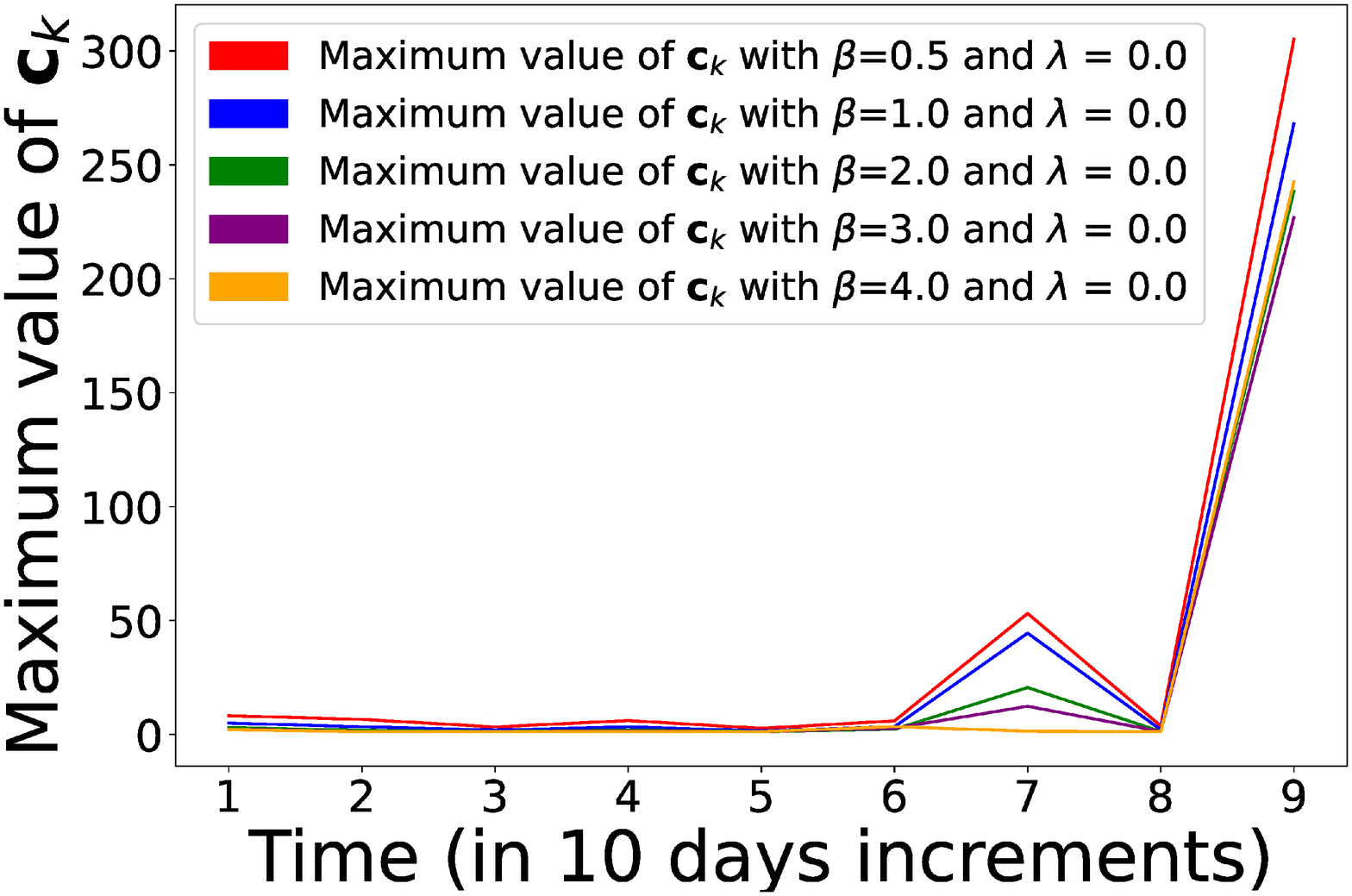}&
		\includegraphics[width=0.31\linewidth]{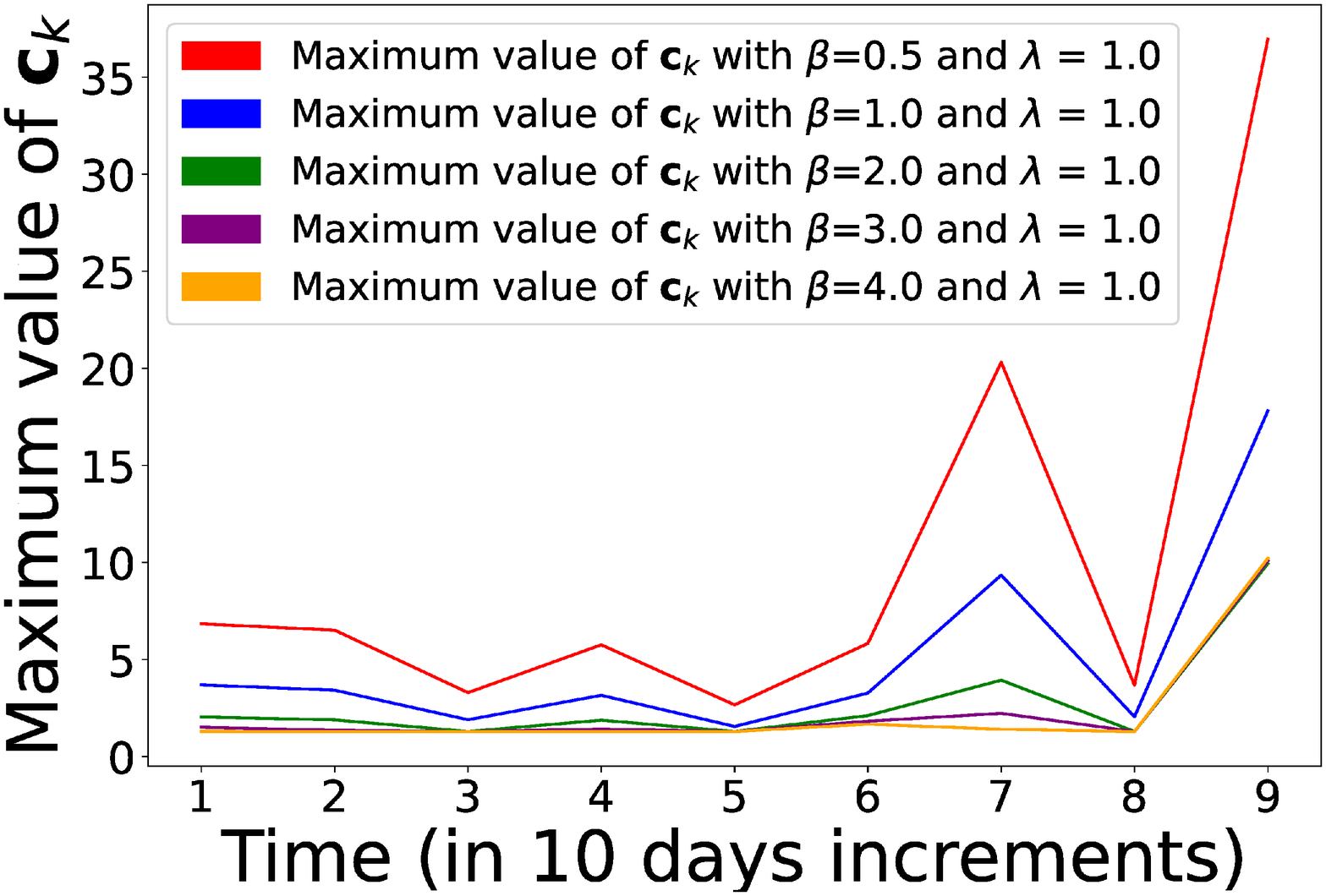}&
		\includegraphics[width=0.31\linewidth]{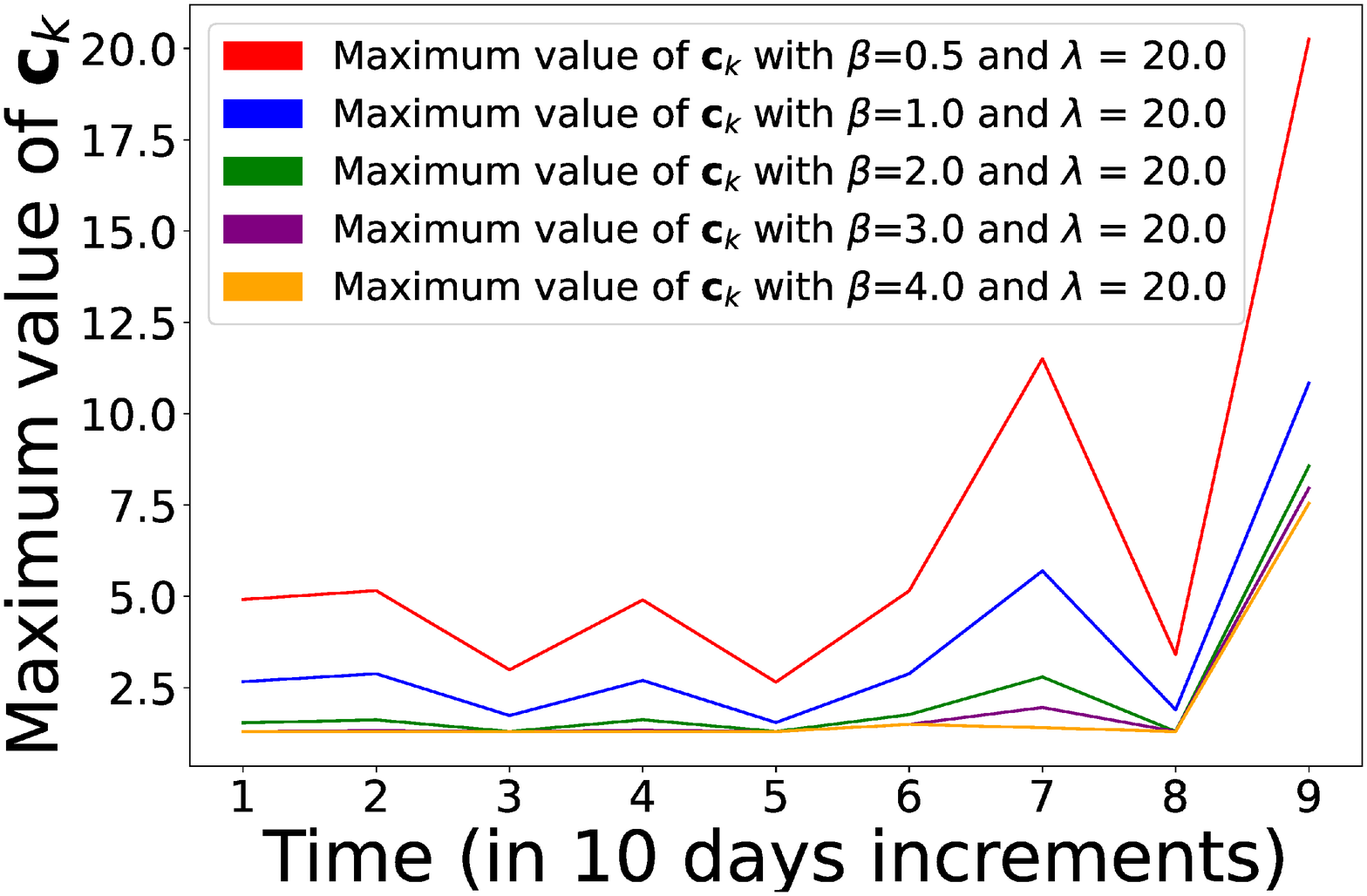}\\
		\includegraphics[width=0.31\linewidth]{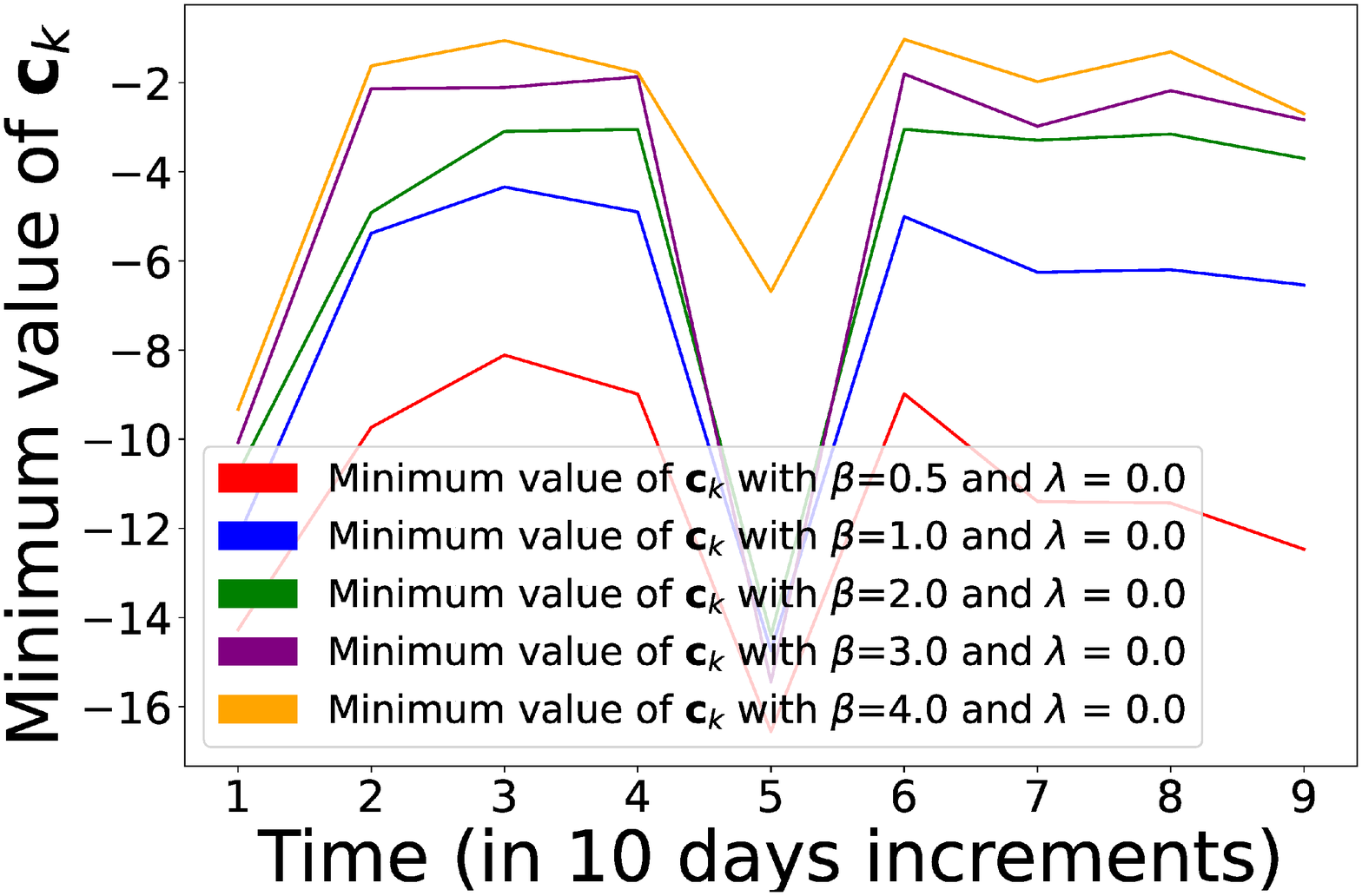}&
		\includegraphics[width=0.31\linewidth]{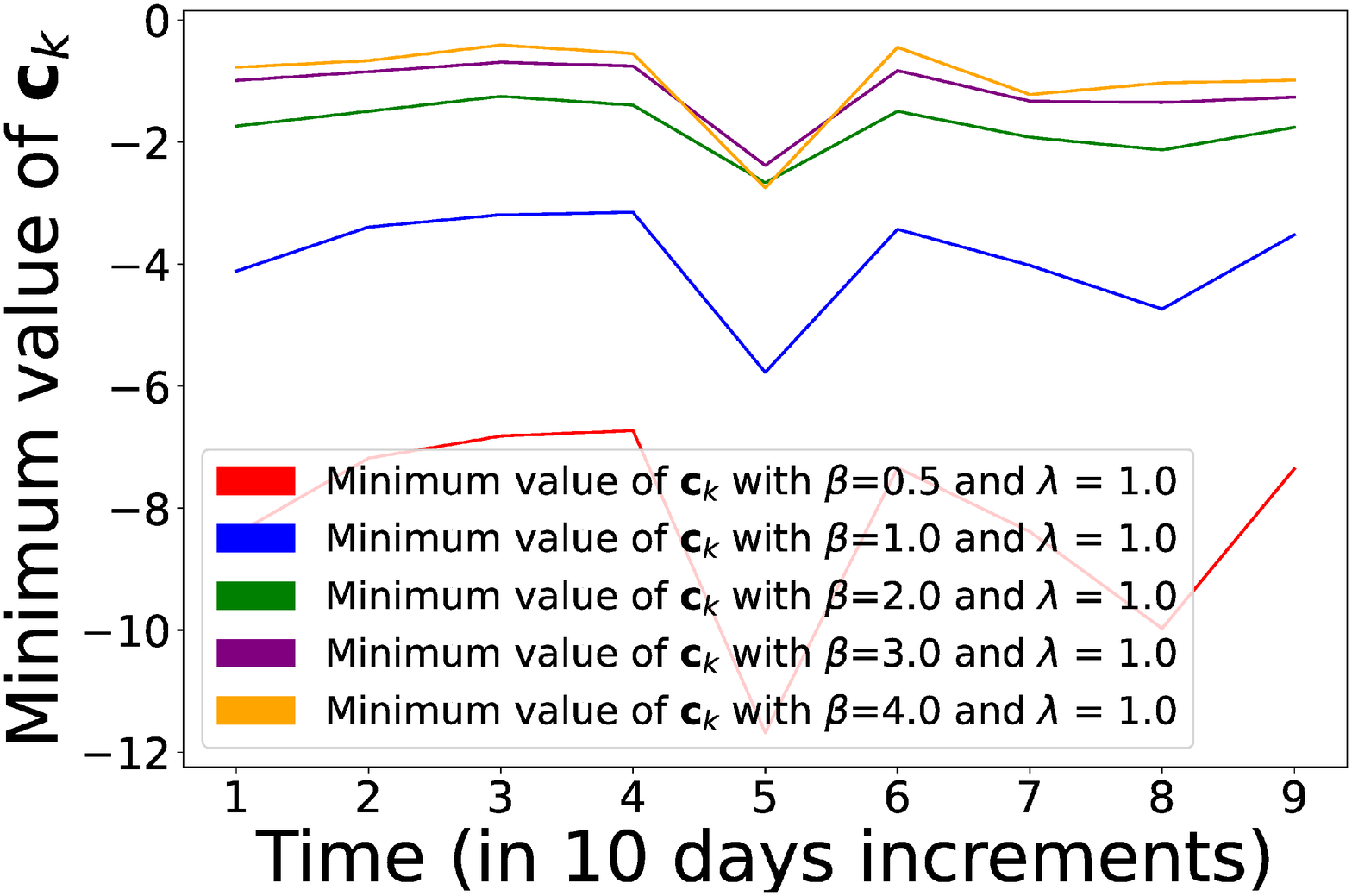}&
		\includegraphics[width=0.31\linewidth]{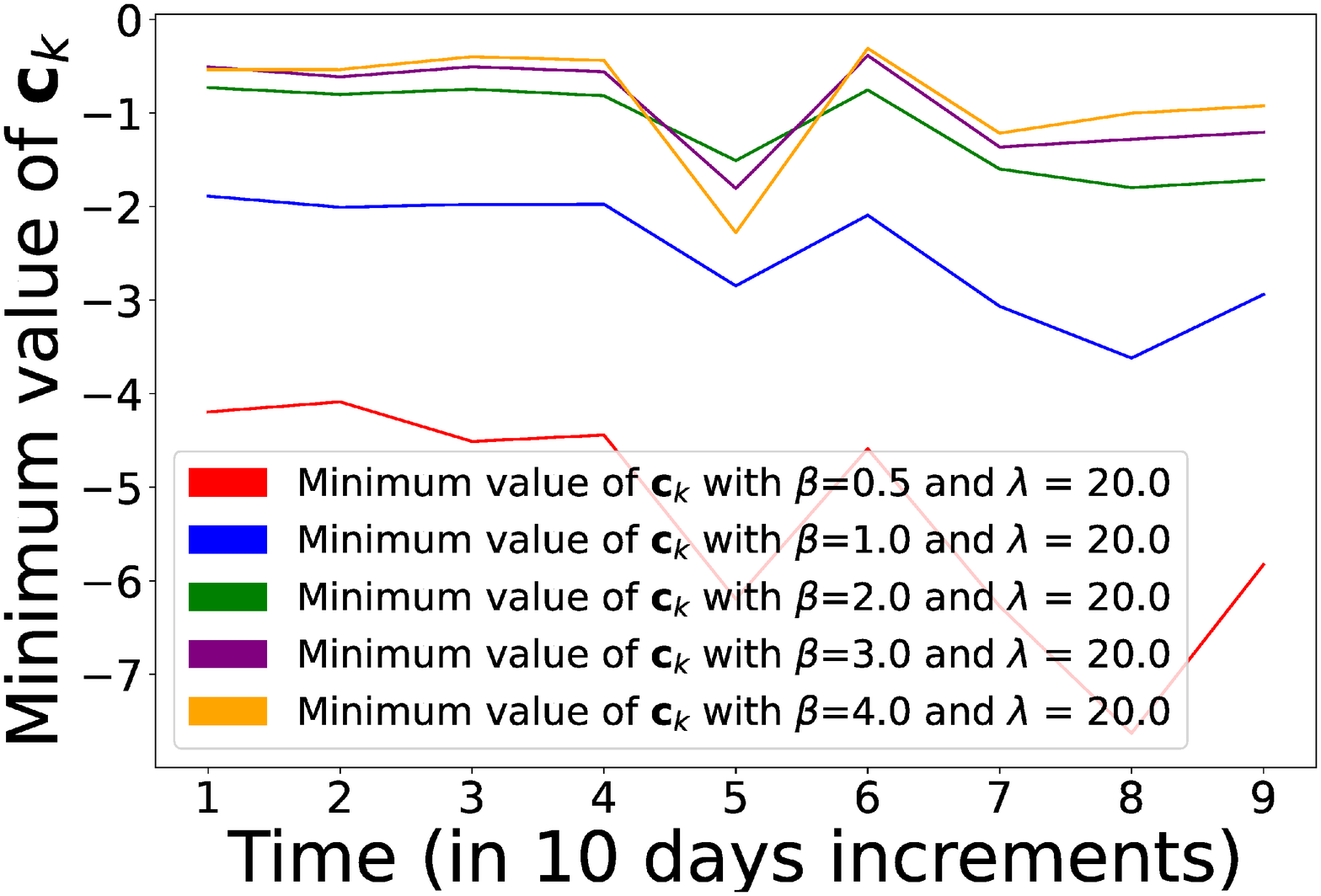}
	\end{tabular}
	\caption{This figure shows how the regularisation minimised the maximum and minimum values of $\bv{c}_k$, hence it shrunk the range of $\bv{c}_k$ values. Without regularisation ($\lambda = 0$) certain $c$ values are too large. }
	\label{fig:reg_on_c}
\end{figure}

The regularization parameter $\lambda$ (recall equations (\ref{eqn:mse_reg}), (\ref{eqn:tichonov}) and (\ref{eqn:singleerror})) is used to minimise the differences in behaviour (that is, parameters $c_k(\bv{r})$ and $u(\bv{r},10k)$) between neighbouring UTLAs. These differences are a measure of model complexity. We illustrate the effect of $\lambda$ on our fitting results. Figure \ref{fig:reg_neigh} shows that increasing $\lambda$ reduces the differences in the total variance of $u$ taken over $K \in \{5, 10\}$ nearest neighbours for all UTLAs. For  $0.05<\beta \leq 2$ the regularisation has a stronger effect on reducing the total variance in behaviour $u$. 
In Figure \ref{fig:reg_scores_lam} we show how the regularisation parameter affects the fitting score $MSE_{\text{reg}}$ (see Equation (\ref{eqn:mse_reg})). 
\noindent
For $\lambda<7$ better fitting scores are achieved with higher values of $\beta$, that is, $\beta \in \{3, 4\}$. At $\lambda \approx 7$ the fitting scores for all studied $\beta$-s are approximately equal, and the trend changes for $\lambda > 7$ so that for $\beta < 3$ the fitting score is better than for $\beta \in \{3, 4\}$. Figure \ref{fig:reg_on_c} shows how $\lambda$ reduces the range of behavioural parameter values, $\bv{c}_k$.

\begin{figure}%[tbhp]
	\centering
	\begin{tabular}{cc}
		\includegraphics[width=0.5\linewidth]{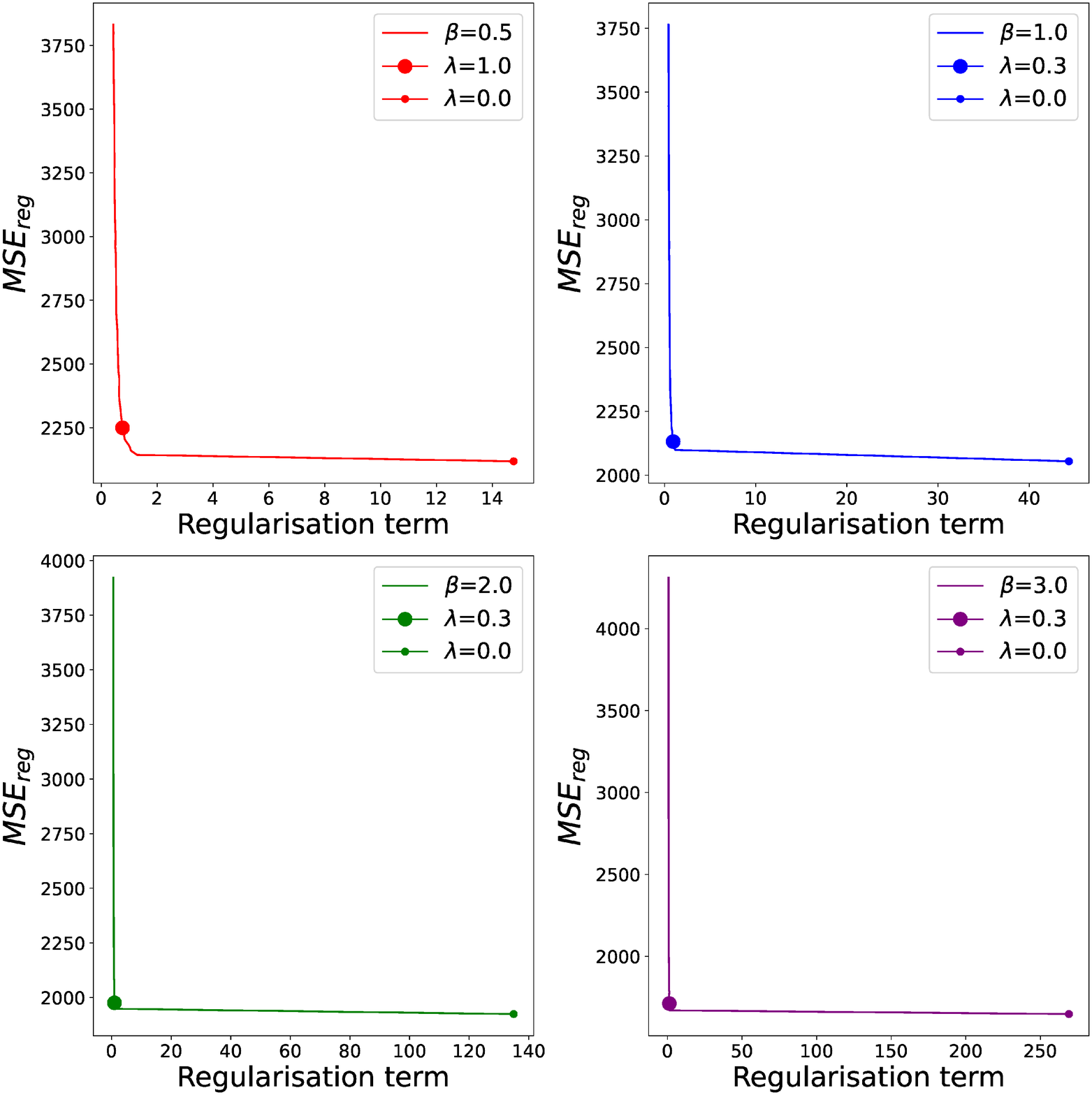}
		&\includegraphics[width=0.5\linewidth]{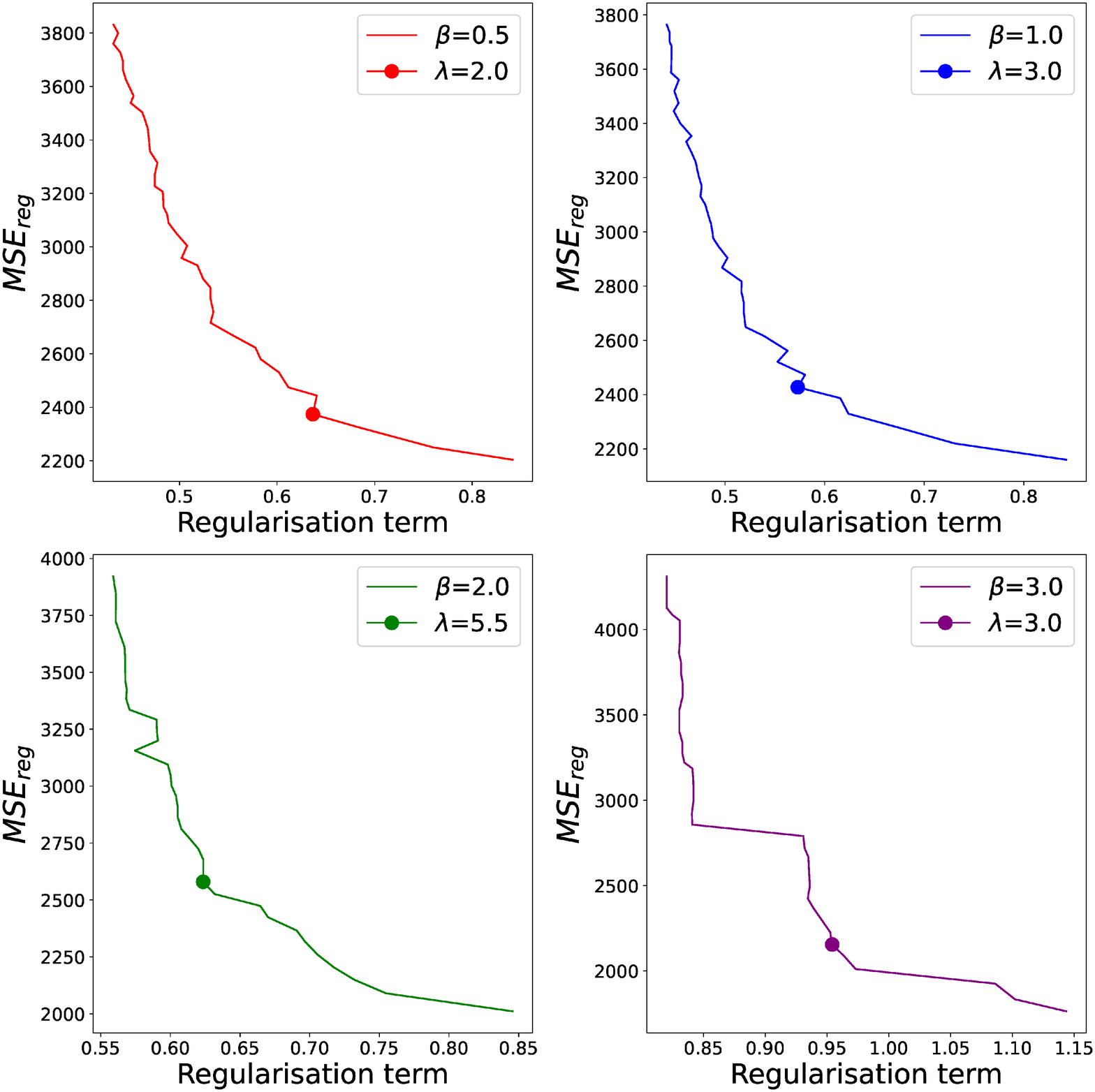}
	\end{tabular}
	\caption{Plots of the average regularisation term values $\frac{1}{n-1} \beta^2\sum_{k=1}^{n-1} ||W\bv{c}_k  - \bv{c}_k ||_2^2$ against $MSE_{\text{reg}}$ score. In the LHS figure we included the case with no regularisation which shows a jump in size of $\frac{1}{n-1} \beta^2\sum_{k=1}^{n-1} ||W\bv{c}_k - \bv{c}_k ||_2^2$ value. The RHS figure excludes the case when $\lambda = 0$ to demonstrate the effect of the regularisation on the $MSE_{\text{reg}}$ score. In the same plot the found elbow/knee points may serve as the candidates for optimal $\lambda$ values.}
	\label{fig:knee_plots}
\end{figure}

\begin{figure}[h!]
	\centering
	\begin{tabular}{ll}
		\includegraphics[width=0.5\linewidth]{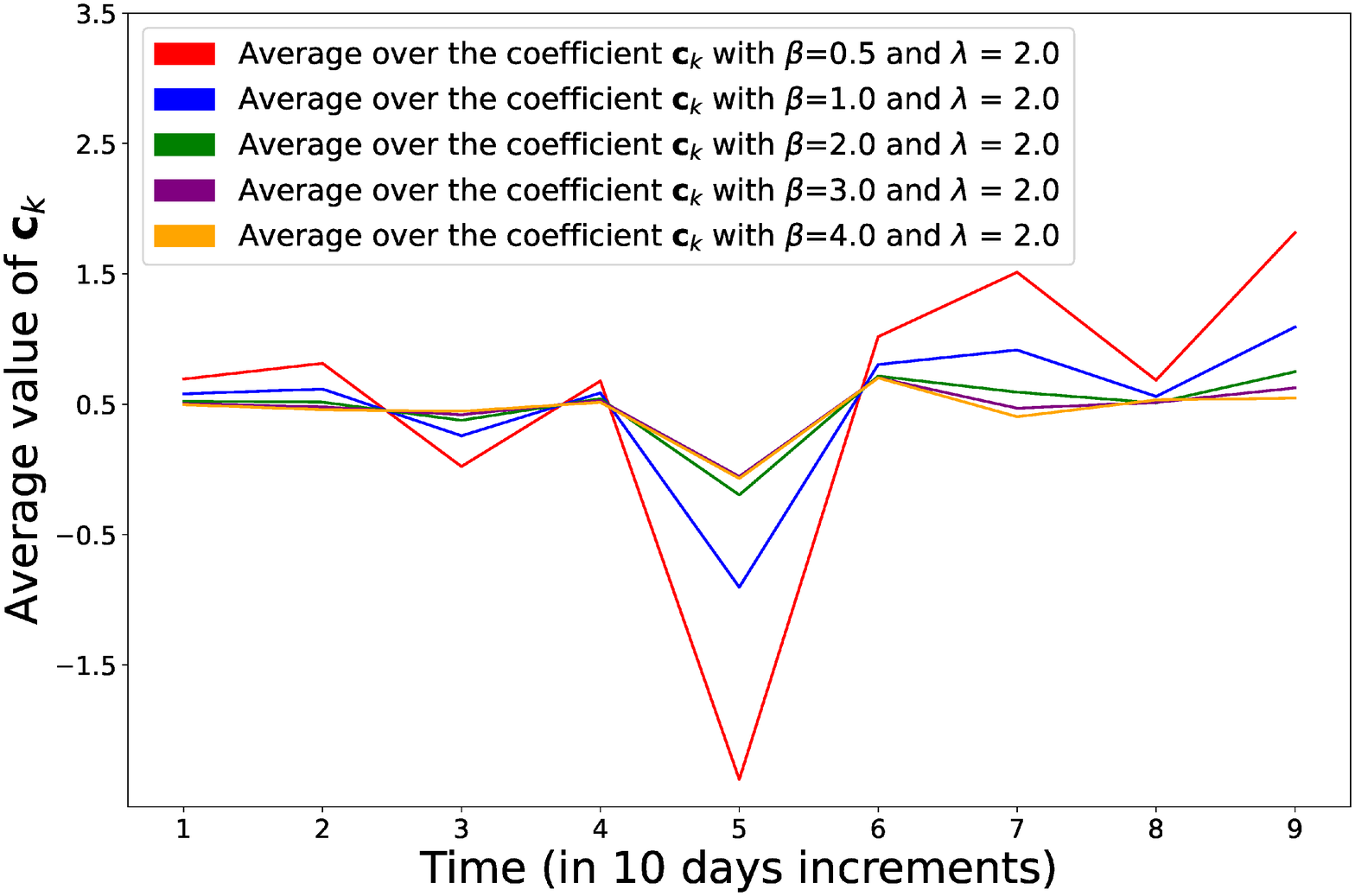} & \includegraphics[width=0.5\linewidth]{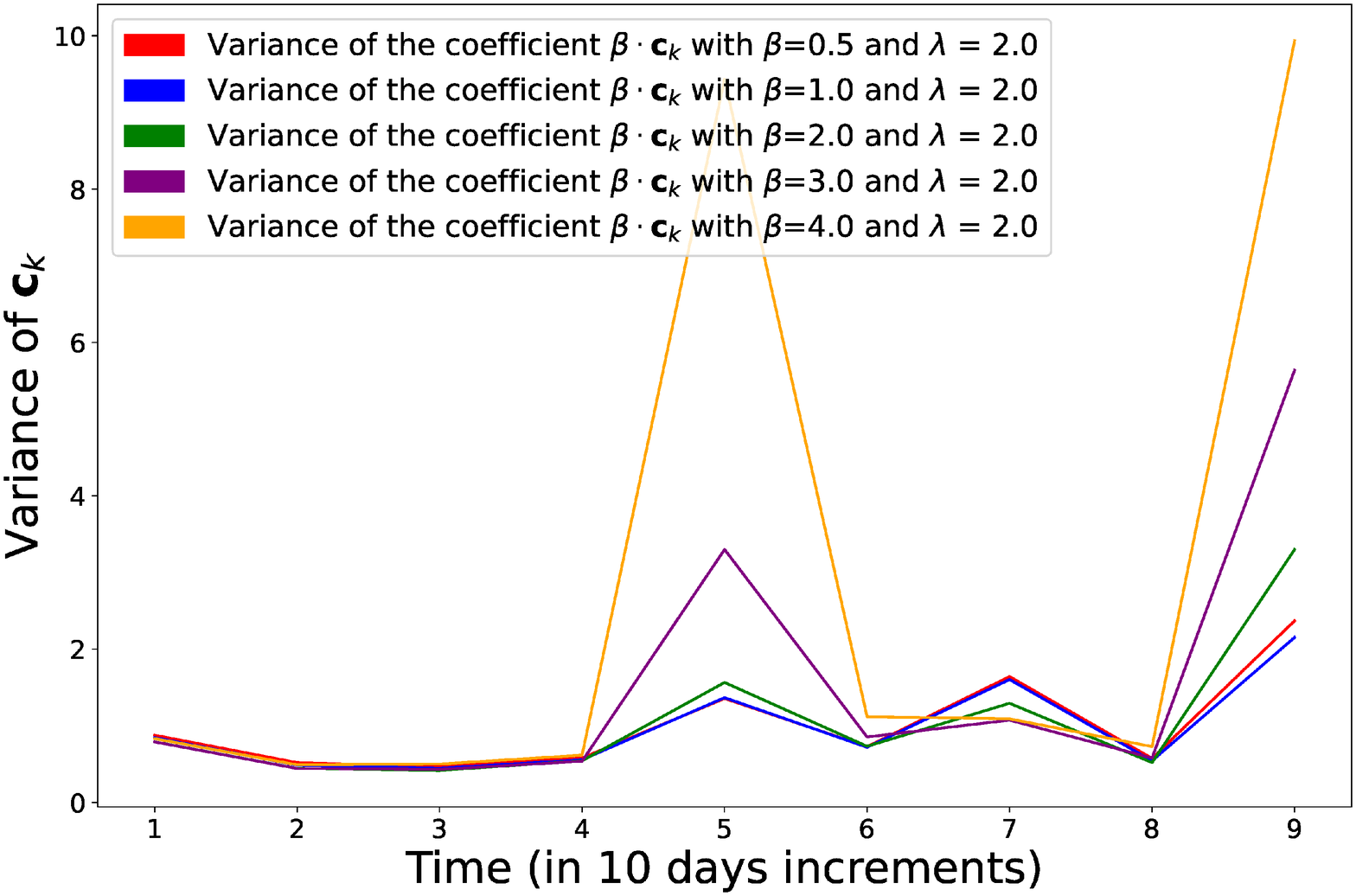}
	\end{tabular}
	\caption{The average value of calibrated parameter $\bv{c}_k$ and the variance of $\beta \cdot \bv{c}_k$ for different values of $\beta$ with regularisation $\lambda = 2$.}
	\label{fig:betas}
\end{figure}

The above mentioned figures demonstrate that regularisation reduces fluctuations in behavioural parameters $\bv{c}_k$ and $u$. These fluctuations are a form of additional model complexity, and should be reduced in order to provide the simplest possible explanation of observed behaviour (Occam's razor \cite{mac03}). From an intuitive perspective, large concurrent fluctuations in behaviour between neighbouring cells appear anomalous. For every $\beta$ value tested, we use the ``knee method'' \cite{sat11}  to determine the optimal $\lambda$ which balances model complexity (as measured by the regularization term) against model fit ($MSE_{\text{reg}}$), as shown in Figure \ref{fig:knee_plots}.  In Figure \ref{fig:betas}, for reference, we show the mean and variance of coefficients $\bv{c}_k$ over different time periods, generated with $\lambda = 2.0$ (the value used in our simulations in section \ref{sec:sim}), for $\beta \in \{0.05, 0.1, 0.5, 1.0, 2, 3, 4\}$.

\subsection*{An approach for finding the optimal value of $\beta$}

\begin{figure}%[tbhp]
	\centering
	\includegraphics[width=0.5\linewidth]{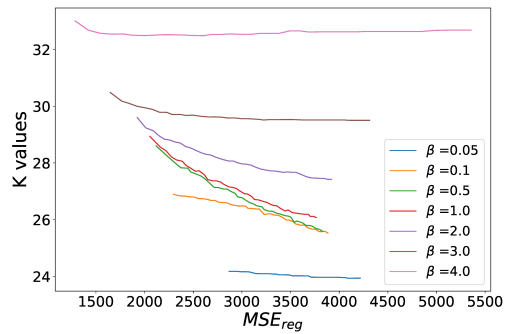}
	\caption{Plots of $MSE_{\text{reg}}$ score against total variance in $u$ over all $5$ nearest UTLAs ($K$ values). Note that $K$ values increase on average as $\beta$ increases. }
	\label{fig:scorecomp}
\end{figure}

\begin{figure}[h!]
	\begin{tabular}{cc}
		\includegraphics[width=0.5\linewidth]{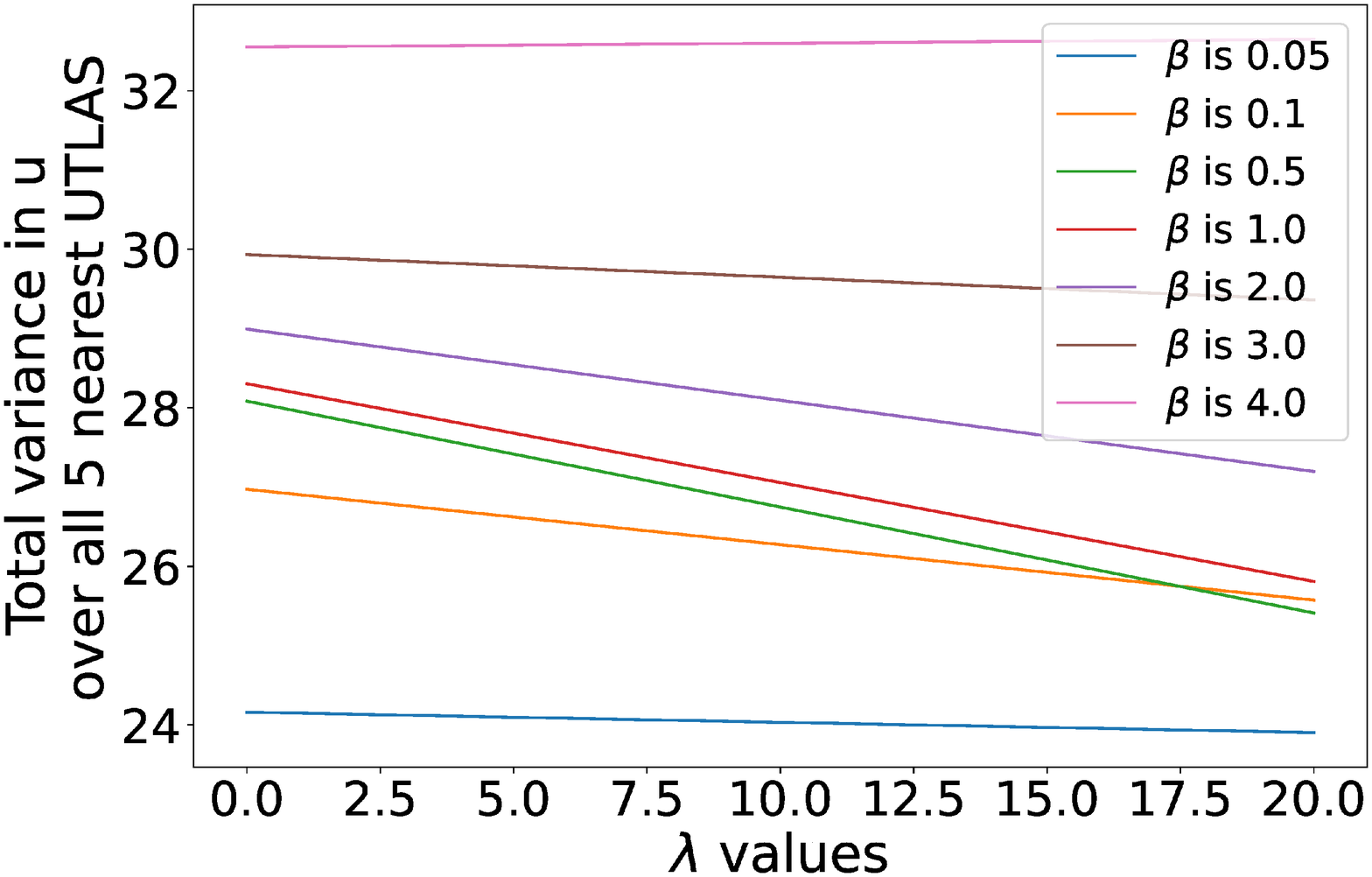} & \includegraphics[width=0.5\linewidth]{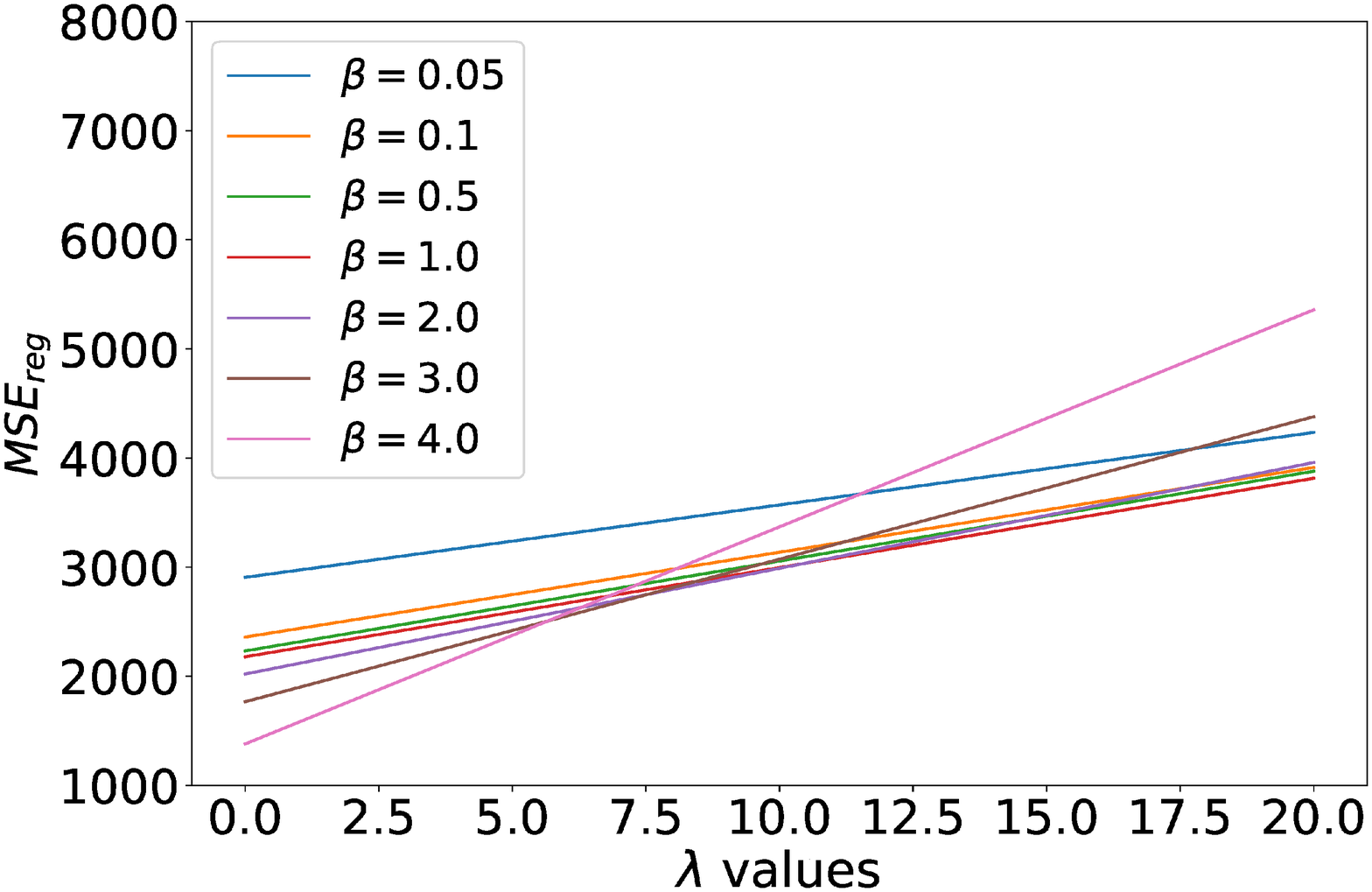} 
	\end{tabular}
	\caption{Least squares line fit to Figures \ref{fig:reg_neigh} (LHS) and \ref{fig:reg_scores_lam} (RHS) }
	\label{fig:str_line}
\end{figure}

\begin{figure}%[tbhp]
	\centering
	\includegraphics[width=0.7\linewidth]{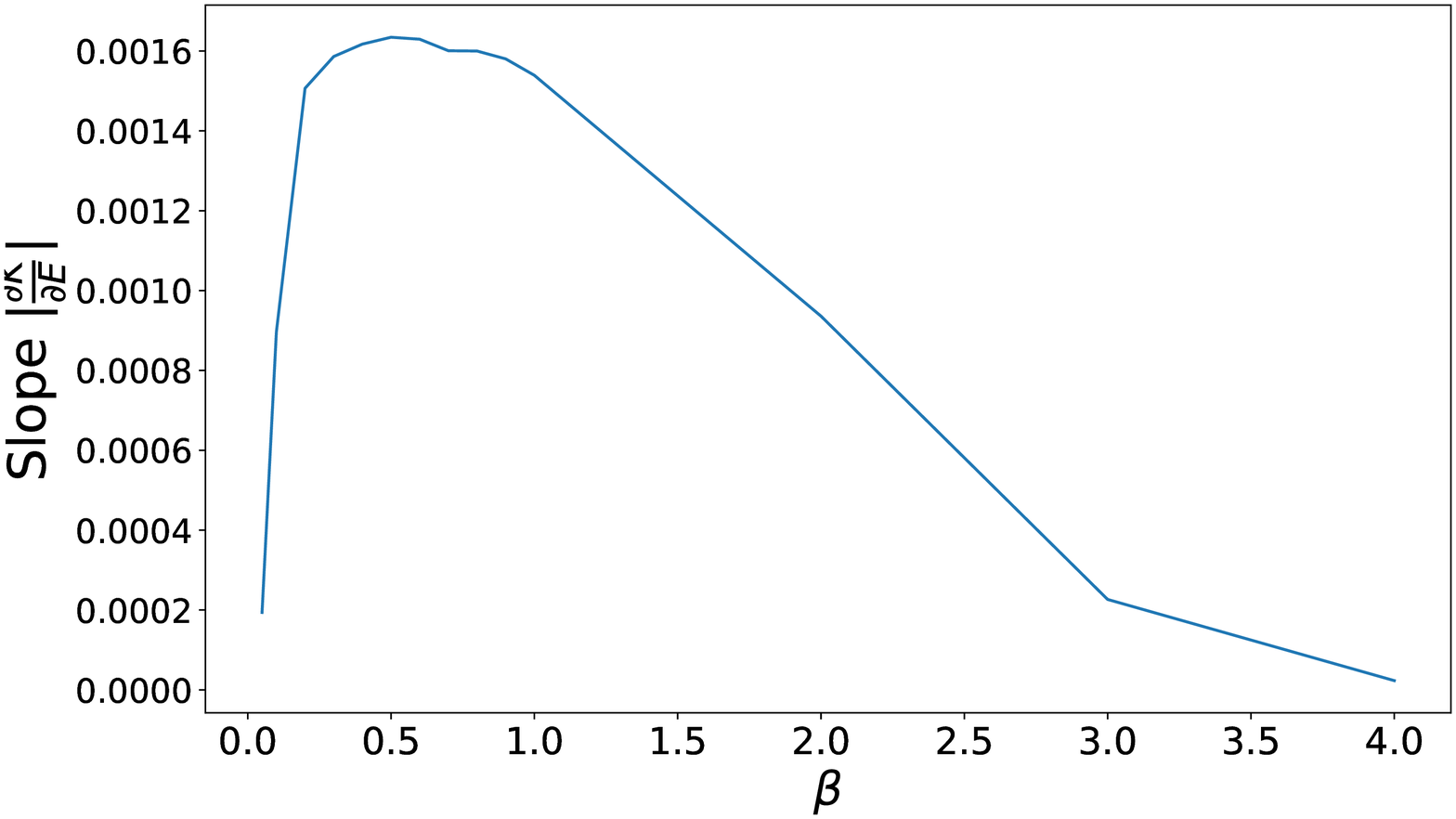}
	\caption{Plot of $\left|\frac{\partial K }{\partial E}(\lambda, \beta)\right|$ for different $\beta$ values. The maximum is at $\beta = 0.5$.}
	\label{fig:slopes}
\end{figure}

Let us denote the score $MSE_{\text{reg}}$ (see Figure \ref{fig:reg_scores_lam}) as $E(\lambda, \beta)$ (a function of $\lambda$ and $\beta$), and the total variance of $5$ nearest neighbours (see Figure \ref{fig:reg_neigh}) as $K(\lambda, \beta)$.
To find the optimal $\beta$ we need to analyse the graphs of $K \equiv K(\lambda, \beta)$ as a function of $E \equiv E(\lambda, \beta)$ for $\beta \in (0,4]$ (see Figure \ref{fig:scorecomp}).
The relationships in Figures \ref{fig:reg_scores_lam} and \ref{fig:reg_neigh} are approximately linear allowing us to approximate them with least-squares linear regression as follows $E(\lambda, \beta) \approx m_{\beta} \lambda + c_\beta $ and  $K(\lambda, \beta) \approx m^{\prime}_\beta \lambda + c_\beta^{\prime} $ (see Figure \ref{fig:str_line}) for some real numbers $m_{\beta}$, $m_{\beta}^{\prime}$, $c_{\beta}$ and $c_{\beta}^{\prime}$ depending on the value of $\beta$.
\noindent 
We have studied the slopes from Figure \ref{fig:scorecomp}, 
in particular, we have that 
\begin{align*}
\frac{\partial K}{\partial E} = \frac{\frac{\partial K}{\partial \lambda}}{\frac{\partial E}{\partial \lambda}} \approx \frac{m_\beta^{\prime}}{m_\beta}. 
\end{align*}
This measures the rate of change of complexity with respect to accuracy. 
%Note that at the boundary of our set of $\beta$ values, that is, $\beta \in \{0.05, 4\}$ the slopes are close to zero, indicating that the regularisation hardly affects the variability in behaviour $u$ among nearest neighbours. Lower $\beta$ values give less accurate fit, but have lower complexity. 
In order to find a balanced solution (between the fit and complexity) we analyse the magnitude of $\frac{\pa K}{\pa E}$, as in Figure \ref{fig:scorecomp}. As all the functions in Figure \ref{fig:scorecomp} are decreasing, having a higher magnitude of a slope ($\frac{\pa K}{\pa E}$) indicates that the fitting score is not much affected by regularisation, but the complexity (local variability in behaviour $u$) is quite sensitive to the regularisation. Based on that observation, the proposed value for the optimal $\beta$ should have the highest value of the magnitude of the slope. We find this to be $\beta = 0.5$ (see Figure \ref{fig:slopes}). The proposed value of $\beta$ is a suggestion based on observations which depend on the data set used in this research.

\subsection*{Difference maps}
\begin{figure}[h!]
	\centering
	\includegraphics[width=0.9\linewidth]{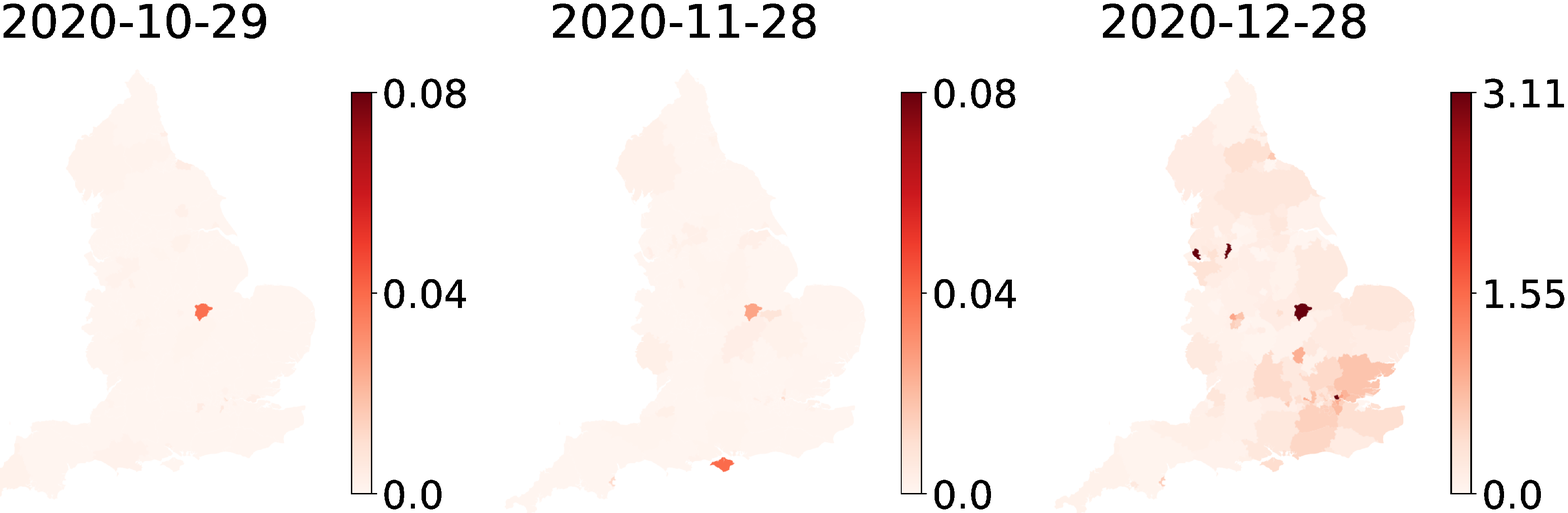}
	\caption{The absolute difference between the maps of active Covid-19 cases (Figure \ref{fig:maps} (a)) and the map of the results from our simulations (Figure \ref{fig:maps} (b)).  }
	\label{fig:diffmaps}
\end{figure}

The absolute difference between the maps in Figure \ref{fig:maps}, that is, 
the map of active Covid-19 cases from the data and the map of cases from the simulations
is captured by Figure \ref{fig:diffmaps}. The corresponding $MSE_{\text{reg}}$ score is $2374.14$, which is very low taking into account the number of total infections over $9$ time periods (see also Figure \ref{fig:overallinf}). 



\bibliography{BSIRS_refs}

\end{document}